\documentclass[superscriptaddress,nofootinbib,pre,twocolumn,floatfix]{revtex4}

\newcommand{\Past}	{ X^{-}}

\newcommand{\Prob}	{ {\rm P} }

\newcommand{\localpast}			{ l^{-}}
\newcommand{\localpastprime}		{ \lambda}

\newcommand{\localfuture}		{ {l}^{+}}

\newcommand{\LocalState}		{ S}
\newcommand{\localstate}		{ s}

\usepackage{amsfonts,amssymb,amsmath,graphicx,hyperref}

\begin{document}
\title{Automatic Filters for the Detection of Coherent Structure in
  Spatiotemporal Systems}
% Or: Our New Filtering Techniques Are Unstoppable!

\author{Cosma Rohilla Shalizi}
\affiliation{Center for the Study of Complex Systems, University of Michigan,
  Ann Arbor, MI 48109 USA}
\email{cshalizi@umich.edu}
\thanks{supported by a grant from the James S. McDonnell Foundation. Present
  address: Statistics Department, Carnegie Mellon University, Pittsburgh, PA
  15213 USA, \texttt{cshalizi@stat.cmu.edu}}

\author{Robert Haslinger}
\affiliation{Martinos Center for Biomedical Imaging, Massachusetts General
  Hospital, Charlestown, MA 02129 USA}
\email{robhh@nmr.mgh.harvard.edu}
\thanks{supported by NIH grants R01 DA015644, R01 MH59733 and the Jenny Fund at
  the Dept.\ of Anesthesia and Critical Care, Massachusetts General Hospital}

\author{Jean-Baptiste Rouquier}
\affiliation{Laboratoire de l'Informatique du Parall\'elisme, \'Ecole Normale
  Sup\'erieure de Lyon, 46 All\'ee d'Italie, 69364 Lyon, France}
\email{jean-baptiste.rouquier@ens-lyon.fr}
\thanks{supported by the ENS-Lyon training program}

\author{Kristina Lisa Klinkner}
\affiliation{Statistics Department, University of Michigan, Ann Arbor, MI
  48109 USA}
\email{kshalizi@umich.edu}
\thanks{supported by a grant from the NSA. After 1 September: Statistics
  Department, Carnegie-Mellon University, Pittsburgh, PA 15213 USA}

\author{Cristopher Moore}
\affiliation{Department of Computer Science, University of New Mexico,
  Albuquerque, NM 87131 USA}
\affiliation{Department of Physics and Astronomy, University of New Mexico,
  Albuquerque, NM 87131 USA}
\affiliation{Santa Fe Institute, 1399 Hyde Park Road, Santa Fe, NM 87501 USA}
\email{moore@santafe.edu}
\thanks{supported by NSF grant PHY-0200909}

%% Begun 26 July 2004
\date{27 July 2005}

\begin{abstract}
Most current methods for identifying coherent structures in spatially-extended
systems rely on prior information about the form which those structures take.
Here we present two new approaches to {\em automatically} filter the changing
configurations of spatial dynamical systems and extract coherent structures.
One, {\em local sensitivity} filtering, is a modification of the local Lyapunov
exponent approach suitable to cellular automata and other discrete spatial
systems.  The other, {\em local statistical complexity} filtering, calculates
the amount of information needed for optimal prediction of the system's
behavior in the vicinity of a given point.  By examining the changing
spatiotemporal distributions of these quantities, we can find the coherent
structures in a variety of pattern-forming cellular automata, without needing
to guess or postulate the form of that structure.  We apply both filters to
elementary and cyclical cellular automata (ECA and CCA) and find that they
readily identify particles, domains and other more complicated structures.  We
compare the results from ECA with earlier ones based upon the theory of formal
languages, and the results from CCA with a more traditional approach based on
an order parameter and free energy.  While sensitivity and statistical
complexity are equally adept at uncovering structure, they are based on
different system properties (dynamical and probabilistic, respectively), and
provide complementary information.
\end{abstract}

\maketitle

\section{Introduction}

Coherent structures are ubiquitous in nature, the result of complex patterns of
interaction between simple units \cite{Ball-tapestry,Cross-Hohenberg}.  Such
structures are not simply epiphenomena of the microscopic physics, but instead
often govern the system's macroscopic properties, providing powerful collective
degrees of freedom---they make good ``handles'' on the system
\cite{Krieger-doing-physics}.  Condensed matter physics provides a host of
equilibrium systems in which this is true \cite{Chaikin-Lubensky}, for instance
in antiferromagnets \cite{Moore-Nordahl-Minar-Shalizi} and liquid crystals
\cite{de-Gennes-Prost-liquid-crystals}, and the anomalous properties of the
high temperature superconductors may be due, in part, to the presence of
ordered charge density waves (stripes) \cite{Kivelson-et-al-in-RMP}.  Coherent
structures also govern the behavior of many non-equilibrium systems
\cite{Cross-Hohenberg}.  The obvious examples are biological systems, which are
rife with coherent structure, and are most certainly far from equilibrium
(until they die) \cite{Winfree-geometry,Camazine-et-al-self-org-in-bio}.
Another example may be found in economics where the spatial pattern of economic
activity has been postulated to be governed by the emergence of coherent
structures in the form of cities and hierarchies of urban centers
\cite{Fujita-Krugman-Venables}.  It has even been proposed that coherent
structures can perform adaptive information processing tasks, and are therefore
the typical physical embodiment of emergent computation
\cite{Steiglitz-Kamal-Watson,Forrest-emergent-computation,Squier-Steiglitz,
  Adamatzky-collision-based, JPC-MM-PNAS}.  This conjecture is supported by
both simulations of biological morphogenesis
\cite{Rohlf-Bornholdt-morphogenesis-long} and experiments on the adaptive
responses of stomata in plants
\cite{Peak-et-al-emergent-computation-in-plants}.  In view of the intrinsic
scientific interest of complex, spatially-extended systems and the potential
power of coherent structures to describe them concisely, it is important to
develop techniques which can uncover and define coherent structures in both the
equilibrium and non-equilibrium cases.  In this paper, we present two such
filters, applicable to spatially extended discrete systems.

Most existing methods for identifying coherent structures rely on knowing some
details about the structure one is looking for.  {\em Matched filters} are
designed so that signals of certain known character maximize the response
\cite{Gershenfeld-modeling}.  Such filters may match either the structures
themselves (e.g.,
\cite{Ruppert-Felsot-et-al-extraction-of-coherent-structures}), or the
uninteresting details of the background in which the structures are embedded
(e.g., \cite{Gunaratne-Hoffman-Kouri-characterizations,Nathan-Gunaratne}).  The
latter technique has been particularly useful in equilibrium systems with
broken symmetry.  The identification of an appropriate order parameter and an
associated free energy implicitly defines a filter which matches the
background, i.e., the ordered state.  Departures from the ordered state, such
as domain walls, vortices and other topological defects constitute mesoscopic
structure.  While extremely successful, this method requires a lot of trial and
error and some knowledge of the underlying micro-dynamics.  In
far-from-equilibrium systems, much of the progress on detecting coherent
structures has involved the application of regular language theory
\cite{Lewis-Papadimitriou-computation} to one-dimensional deterministic
cellular automata
\cite{Wolfram-computation,Attractor-basin-portrait,Hordijks-rule,Pivato-defect-kinematics}.
Specific filters, adapted to particular cellular automata rules, have been
designed to detect structures dynamically
\cite{Grassberger-diffusion,Grassberger-chaos-and-diffusion,
  Boccara-Nasser-Roger-particle-like,Turbulent-pattern-bases,
  Comp-mech-of-CA-example}.  In these cases as well however, one must know a
lot about the system's dynamics to work out, by hand, a suitable filter.
Furthermore, to apply such notions to cellular automata of higher dimension,
one would have to extend regular language theory to such dimensions, which
presents significant difficulties \cite{Two-D-Patterns}.  Extensions to
stochastic systems would be even more problematic.

The two filters introduced in this paper address these issues.  Both are
``automatic'' filters, in that they require no prior information about the
system's microscopic dynamics or the structures generated.  They can be applied
to both equilibrium and non-equilibrium systems of, in principle, any
dimensionality.  One of them is applicable to systems with stochastic dynamics.
The two filters are, however, based upon very different system properties.  \S
\ref{sec:lle} introduces the {\em local sensitivity}, an adaptation of Lyapunov
exponents to cellular automata, measuring the degree to which small
perturbations can alter the dynamics.  \S \ref{sec:sc} defines the {\em local
  statistical complexity}, which measures the spatiotemporal distribution of
the amount of information needed to optimally predict the system's dynamics.
It is therefore a probabilistic measure, completely compatible with any
underlying stochastic dynamics.  Both methods identify coherent structures by
filtering the system's changing configuration with respect to sensitivity (or
complexity), and tracking the changing spatial distributions of these fields.
We apply each filter to the task of detecting coherent structures in cellular
automata.  Sections \S \ref{sec:lle-eca} and \S \ref{sec:sc-eca} look at
``elementary'' (one-dimensional, binary, range-one) cellular automata (ECA),
where we find that both filters readily identify ECA particles, domains and
domain walls, although they emphasize these structures to differing degrees.
In section \S \ref{sec:cca} we use cyclic cellular automata (CCA), a
self-organizing model of excitable media, to compare the local sensitivity and
statistical complexity with a more traditional order parameter approach.  (We
identify the appropriate order parameter for this system in \S
\ref{sec:cca-op}.)  Our conclusion gives a summary and discusses some open
questions.  Finally, appendices discuss previous attempts to adapt the Lyapunov
exponent to cellular automata (\ref{app:lyap_review}), and provide details on
the construction of the CCA order parameter and effective free energy
(\ref{app:order-parameter}).

\section{Local Sensitivity}
\label{sec:lle}

The first of our filters for coherent structures is based upon a measure of
{\it local sensitivity}: the instability of the discrete cellular automata
field under local perturbation.  Filtering with respect to local sensitivity is
motivated by a desire to distinguish autonomous objects from the rest of the
CA's configuration field.  By ``autonomous objects'' we mean those structures that have
the greatest influence upon the future dynamics of the system.  Perturbing an
autonomous object will affect the future a great deal.  In contrast,
perturbations of the rest of the system, the dependent parts, should be quickly
overridden and ``healed'' by the autonomous dynamics.  The set of autonomous
objects constitutes a high-level model of the system, and knowledge of them
should allow us to infer most of the rest of the dynamics, as well as the
objects' own futures.

\begin{figure}[thbp]
  \includegraphics[width=\columnwidth]{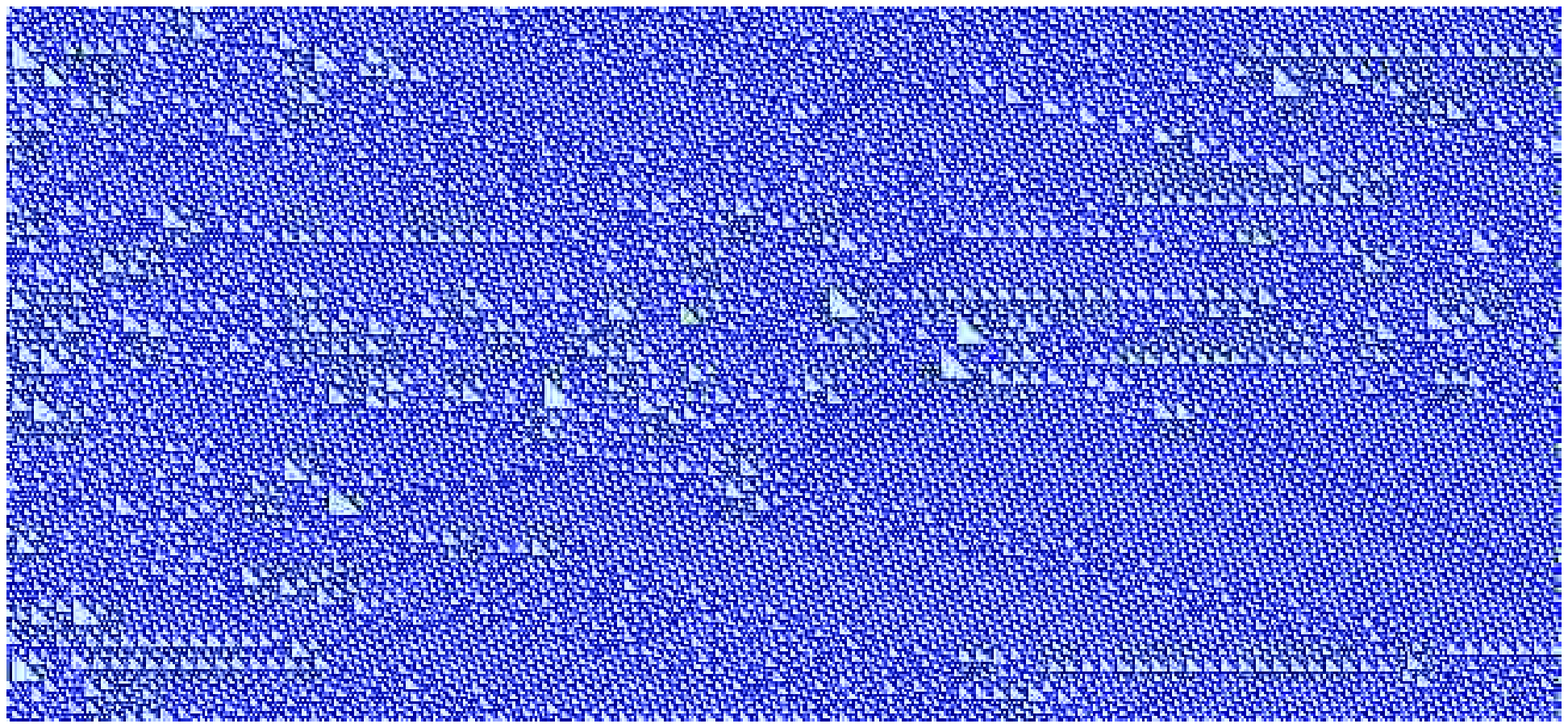}
  \caption{(Color online.)  Typical space-time diagram of rule 110; time
    advances from left to right.  Note the presence of the regular background
    domain, and the particles which stand out in the background by contrast.}
  \label{fig:rule110}
\end{figure}

\begin{figure}[thbp]
\centering
$a$ \includegraphics[height=0.25\textheight]{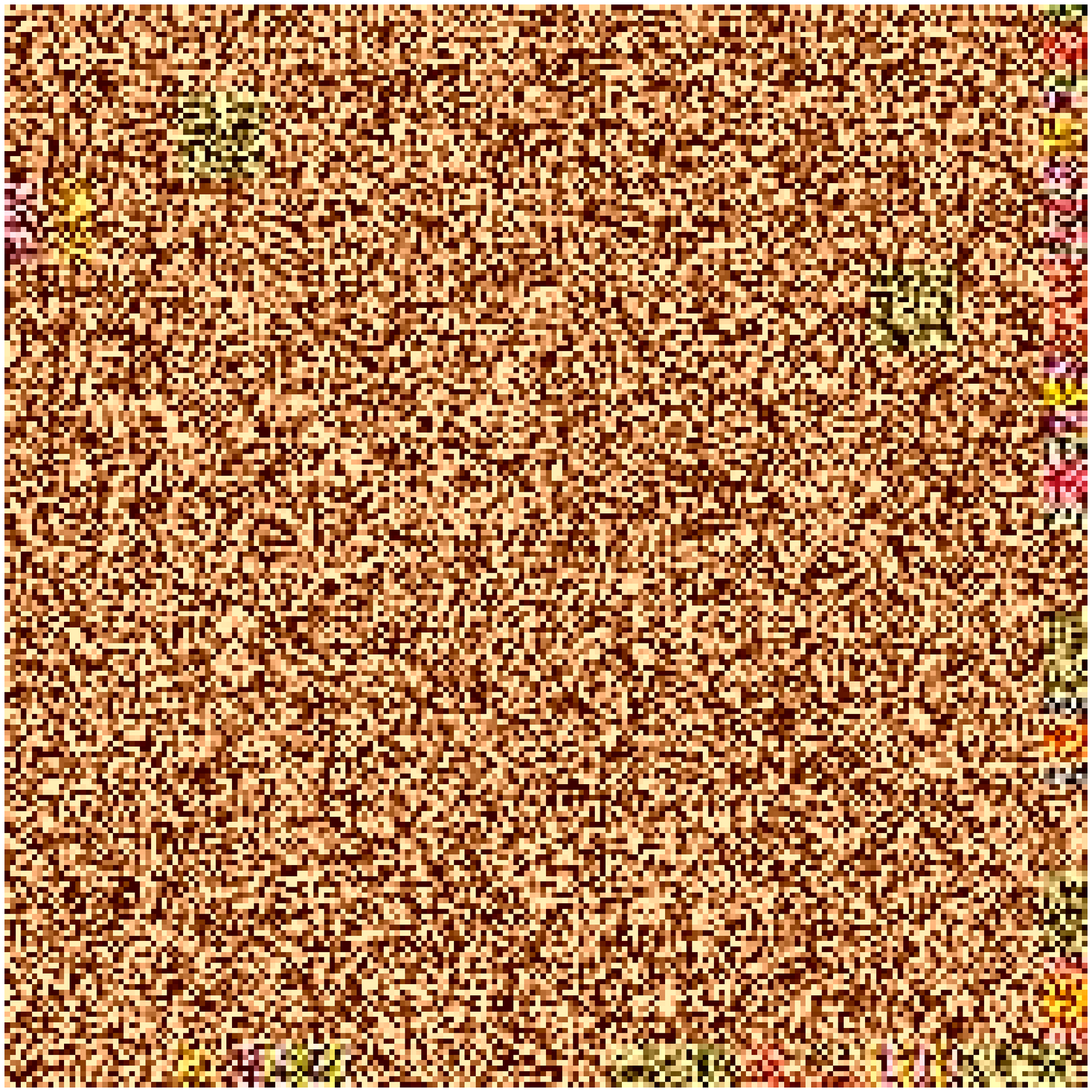}\\
$b$ \includegraphics[height=0.25\textheight]{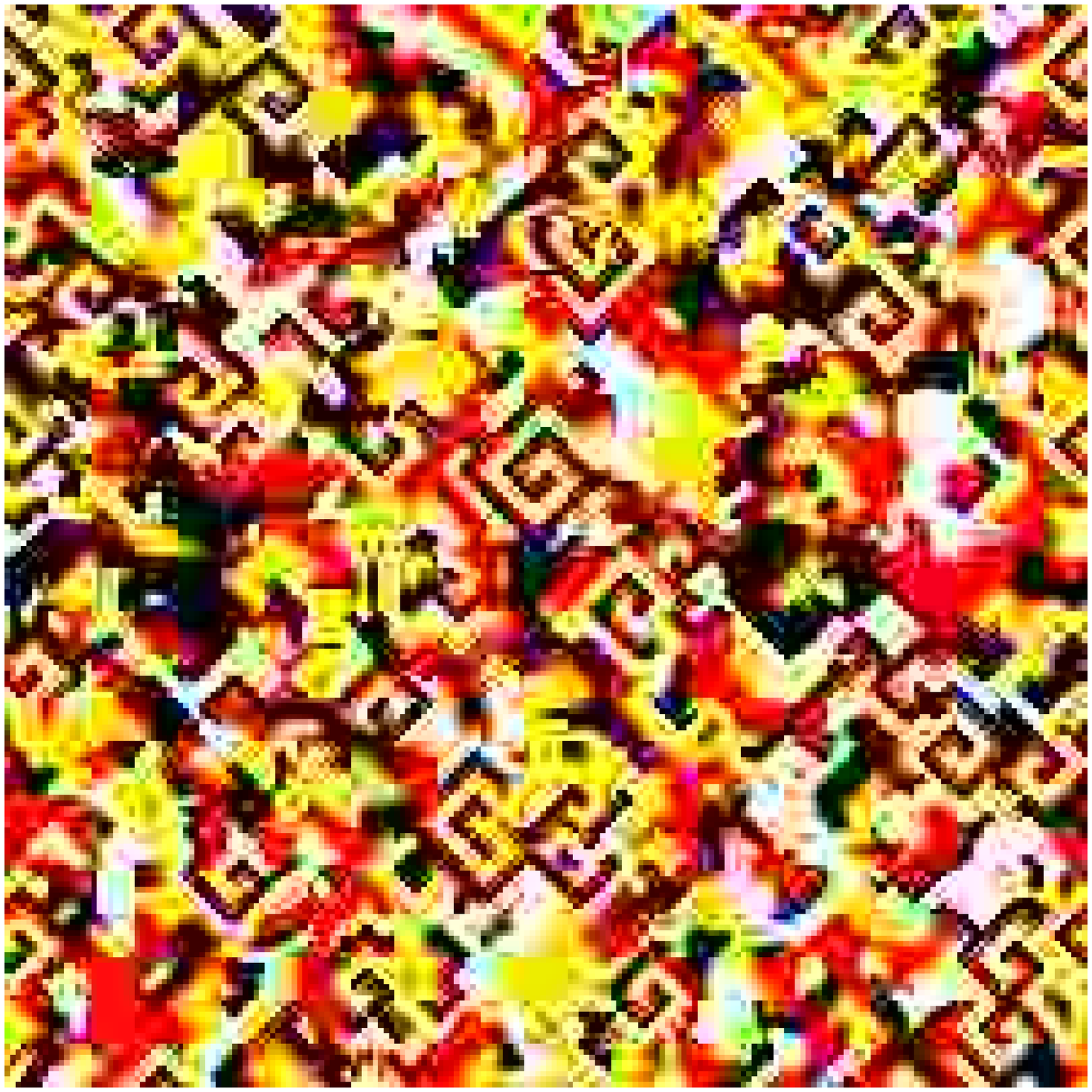}\\
$c$ \includegraphics[height=0.25\textheight]{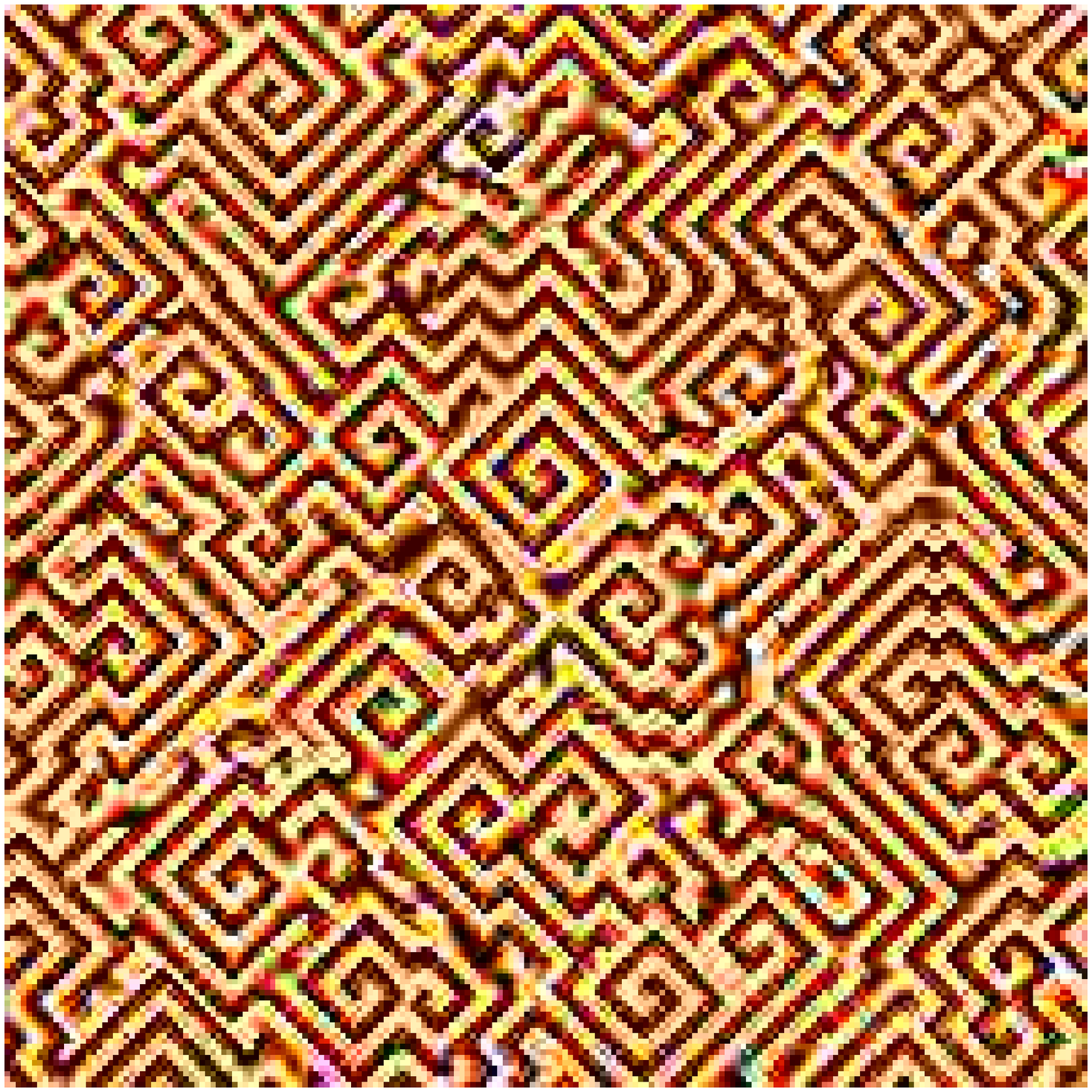}
\caption{(Color online.)  Phenomenology of the cyclic CA (defined in
  \S\ref{sec:cca}) in one of its spiral-forming phases ($\kappa = 4$, $T = 2$,
  $r=1$).  Time progresses from top to bottom, the panels depicting $(a)$ the
  random initial conditions, $(b)$ the formation of spiral wave cores, and
  $(c)$ their eventual domination of the entire lattice.}
\label{fig:CCA-time-evolution}
\end{figure}

A famous example of a system governed by autonomous objects is rule 110 of the
elementary cellular automata.  A rule 110 configuration field (Fig.\ \ref{fig:rule110})
tends to exhibit particles, and knowledge of the position of these particles
carries considerable predictive power
\cite{Cook-on-110-published,Hordijks-rule}.  A second example may be found in
the cyclic cellular automata (see below, \S \ref{sec:cca}).  For certain
parameter values the CCA configuration field evolves into competing spiral waves, the
cores of which are autonomous (see Figure \ref{fig:CCA-time-evolution}).

For these two examples, rule 110 and spiral CCA, the presence of structure is
obvious.  However this is not always the case (see, e.g.,
Ref.\ \cite{Turbulent-pattern-bases}, and the discussion of rule 146 in \S
\ref{sec:lle-eca} below).  Even when it is, reliably identifying all the
structure present in a CA's configuration field is difficult to do by eye and it is
often not at all clear how autonomous such structures are.  One might therefore
hope for an automatic filter capable of not only distinguishing particles,
domains, and extended objects like strings and domain walls in large
spatiotemporal systems, but also inferring their degree of autonomy.

\subsection{Defining Local Sensitivity}

Since we want to filter cellular automata with respect to autonomy, a natural
point of departure is some mathematical formalization of the idea that a
system's future may be very sensitive to details of its present state.
Devaney's definition of chaos \cite{Banks-et-al-on-Devaney}, and the notions of
topological and metric (Kolmogorov-Sinai) entropy rates \cite{Ruelle-Lincei}
are two ways in which this intuition can be formalized.  Here, however, we will
begin with the idea of a dynamical system's Lyapunov exponents, since these
provide a more refined and quantitative measure of sensitivity-dependence than
either the mere fact of chaos, or than the metric entropy rate (which is
generally the sum of the positive Lyapunov exponents).  We briefly review the
definition of Lyapunov exponents, before developing our filter.

A one-dimensional map $f$, for which the system's state is given by a single
real variable $x$, has a single Lyapunov exponent, given by the following
limit:
\[
\lambda = \lim_{n\rightarrow\infty}{\frac{1}{n}\log{\left|D f^n(x_0)\right|}}
\]
where $D$ is the derivative operator.  (The limit can be shown to exist and be
identical for almost all initial points $x_0$.)  The interpretation is that, if
one begins with two points $x_0$ and $x_0+\epsilon$, the distance between their
future images, after $n$ time steps, is approximately $\epsilon e^{n\lambda}$;
this relation becomes exact as $\epsilon \rightarrow 0$.  In the case of a
dynamical system described by $m$ state variables, the Lyapunov exponents are
defined as the $m$ eigenvalues, $\lambda_1 \ldots \lambda_m$, of the matrix
\[
\Lambda = \lim_{n\rightarrow\infty}{{(D^{\dag}f^n(x_0)Df^n(x_0))}^{1/2n}}
\]
where $D$ is again the derivative operator, and $A^{\dag}$ denotes the adjoint
of $A$.  These give the exponential growth rates of small perturbations applied
along different directions (which are not, however, the eigenvectors of
$\Lambda$, but those of $Df$).  The spectrum of Lyapunov exponents thus gives a
fairly detailed picture of the kinds and degrees of sensitivity displayed by
the system dynamics.

These definition take no account of the system's spatial structure; this should
be rectified in an application to cellular automata.  In addition, we must
define the exponents in a way which is spatially {\em local}, so that their
variation over space can be studied, and structures identified.  In continuous,
spatially-extended systems the traditional approach has been to consider the
global configurations of the system as points in a Hilbert space
\cite[p.\ 53]{Ruelle-Lincei}, and then expand the above definition to include
infinite-dimensional spaces.  Although this produces a meaningful set of global
exponents, these exponents contain no information about whether the system is
more sensitive at {\em particular} points in space.  Such information is
crucial for identifying coherent structure.  Also crucial is an ability to
identify structures at varying spatial scales.  For example the elementary CA
rule 110 has particles that can be over twenty cells wide---and all parts of
the particle need not be equally sensitive.  Some more recent work has
attempted to define Lyapunov exponents in a spatially local manner, for
precisely these reasons \cite{Cross-Hohenberg}.

A second challenge is to define Lyapunov exponents in a way which is applicable
to systems which are {\em discretized}, both in state and space.  The classical
definition of the Lyapunov exponent invokes limits taken as two initial
conditions approach arbitrarily close to one another in continuous state
spaces.  (This is implicit in the use of derivatives.)  In cellular automata,
in which the state is discretized, the closest distinct configurations possible
are two configurations differing only on one cell\footnote{There are metrics,
  such as the Cantor metric, in which CA configurations can approach
  arbitrarily close, but they are inappropriate for local studies, such as
  ours.}.  Space is also discretized: there are no distances intermediate
between one and two differing cells.  Several attempts have been made to frame
definitions of Lyapunov exponents which accommodate these facts; we discuss
these attempts, and why we feel they are not suitable for our present purposes
in Appendix \ref{app:lyap_review}.  Here we take the straightforward approach
of perturbing small patches of contiguous cells, evolving the system forward in
time, and measuring the distance between the perturbed and unperturbed
configurations.  This eliminates the need to take derivatives, and, by varying
the size of the perturbed region, one can examine the structures present at
different length scales.

Formally, we define the local sensitivity at point $\vec{r},t_0$ as the result
of the following calculation.  Let $P$ be the set of all cells at distance at
most $p$ of $\vec{r}$.  The parameter $p$ is called the \emph{perturbation
  range}.  Let $S$ be the set of all possible configurations restricted to $P$
(i.e., the set of words of length $|P|$ taking the states as alphabet).  For
one $s\in S$, we set the cells in $P$ to their state in $s$, and keep the
original configuration for all other cells.  Then we let the automaton evolve
for $f$ steps.  We name the parameter $f$ the \emph{future depth}.  Finally, we
measure the area of the difference plumes, the total number of cells which
differ from the original configuration, at each time step and compute the mean,
weighting the area at time $t$ by $1/|w|$, where $w$ is the set of points at
time $t$ that depend on $(\vec{r},t_0)$.  We average this on all $s\in S$ to
get the sensitivity at that point, $\xi(\vec{r},t_0)$, which, intuitively,
gauges how unstable the system is with respect to perturbations there.  Note
that $\xi$ is normalized so that it always lies between $0$ and $1$,
facilitating the comparison of distinct rules.

Two parameters enter into our calculation of $\xi$, though we suppress them in
the notation: the perturbation range $p$ and the future depth $f$.  Both must
be chosen with some care.  If they are too small, the calculated $\xi$ is
excessively sensitive to fluctuations inside the background domain or within
particles, but over-large parameter choices tend to make things blurry, and are
quite time-consuming.  Choosing the right $p$ is like adjusting a microscope to
the size of the object one wants to observe.  If the magnification factor is
too high, one might miss the bigger structures, or even encounter diffraction
artifacts (counterpart of being too close to the discretization scale).  It may
seem self-defeating that we should need to know the size of the relevant
structures to successfully use an automatic filter.  However there may be
structures at different scales, and varying $p$ allows one to filter these
preferentially.  Similarly, the coherent structures of many systems can be
arranged in a causal hierarchy, where higher-level objects drive lower-level
ones, and this can be detected by increasing the future-depth $f$, which tends
to pick out the most autonomous features.  (See the example of ECA rule 146
below, especially Figure \ref{fig:rule146_results}, obtained with $f=30$.)
Note that the levels of the spatial and causal hierarchies need not be aligned
with one another, i.e., the largest structures need not be the most autonomous.

\subsection{Results from Elementary Cellular Automata}
\label{sec:lle-eca}

\begin{figure}[tbhp]
  \centering
$a$ \includegraphics[width=0.95\columnwidth]{rule110_states_step1000_c}  
  \par\smallskip
$b$ \includegraphics[width=0.95\columnwidth]{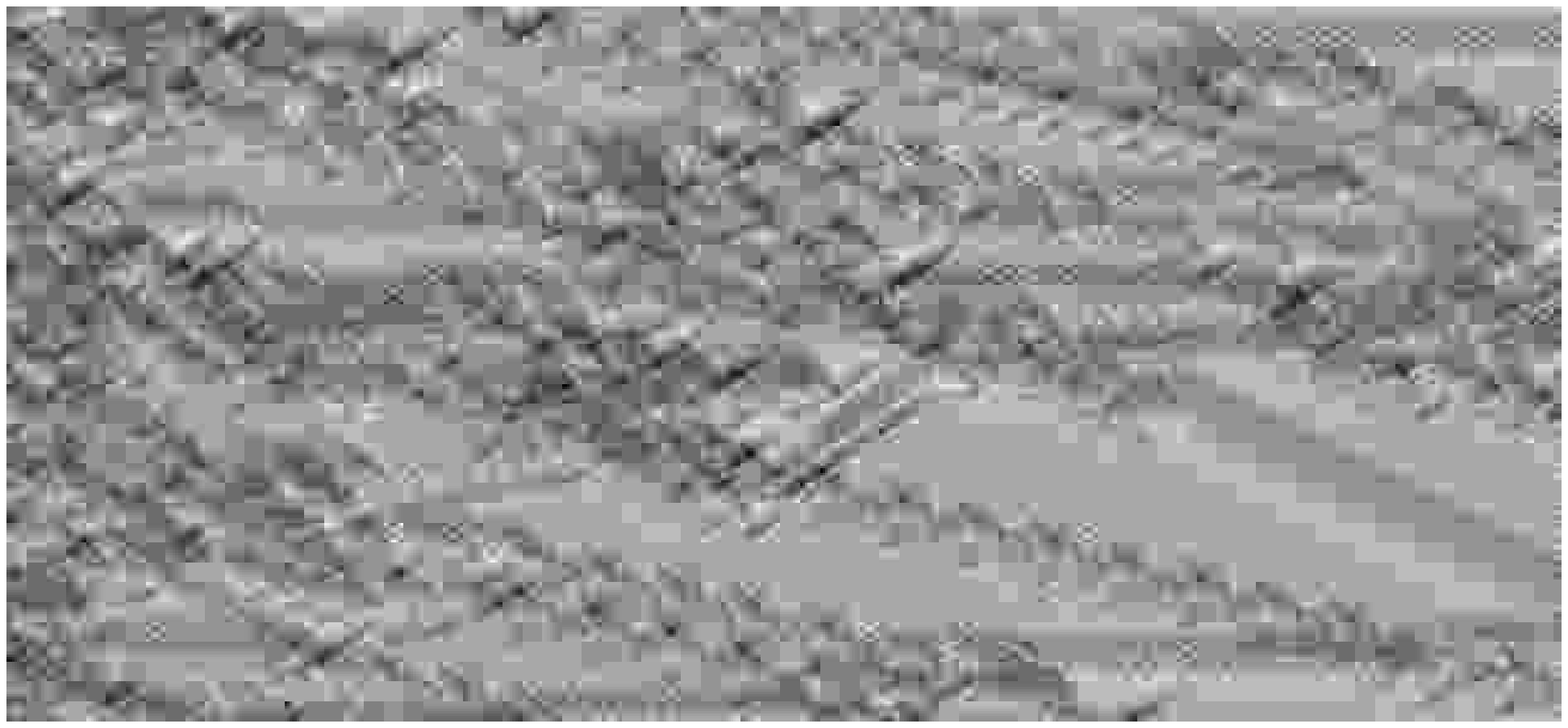}
  \par\smallskip
$c$ \includegraphics[width=0.95\columnwidth]{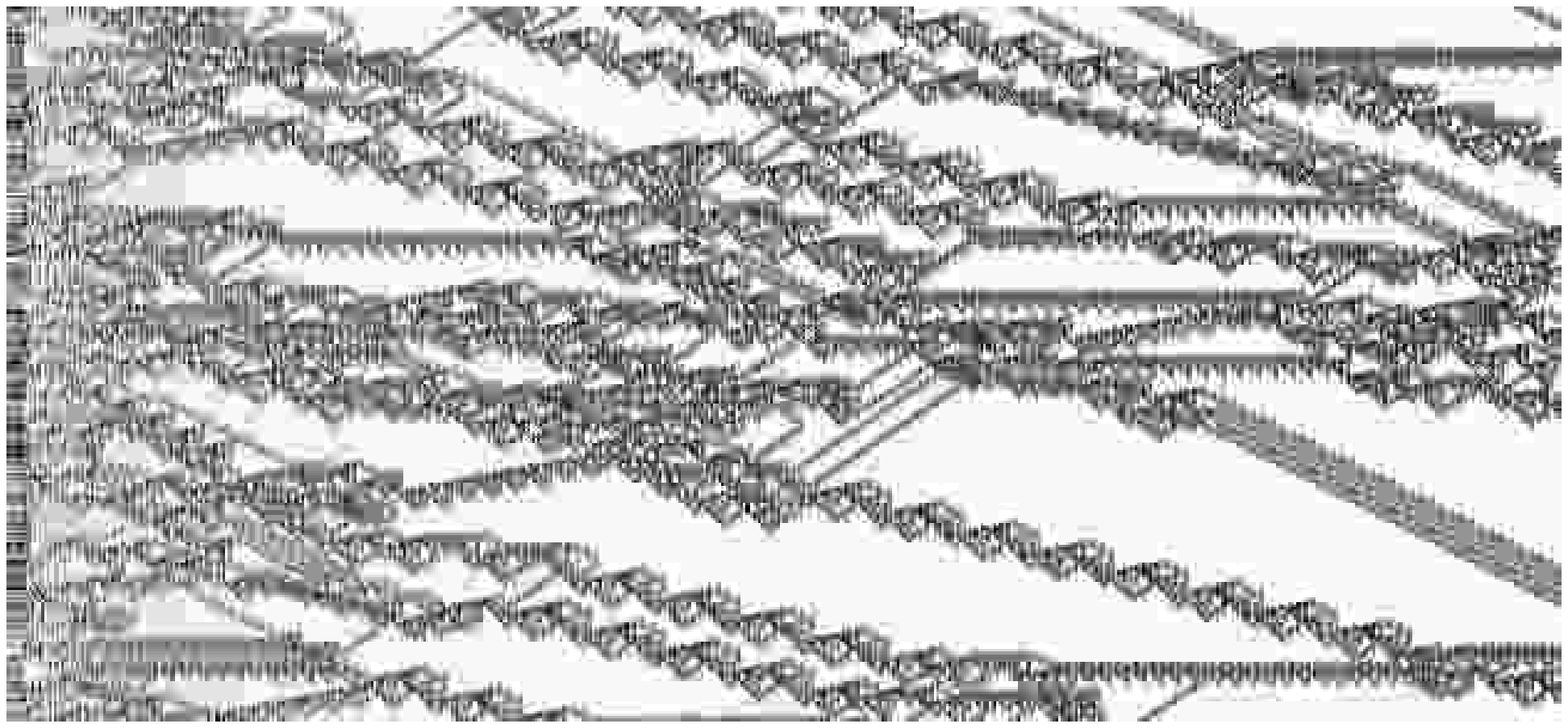}
  \caption{Evolution of rule 110 from a random initial condition (the same as
    that shown in Figure \ref{fig:rule110}, repeated for convenience): $a$,
    configuration field; $b$, sensitivity field (calculated with $p=1,\ f=10$);
    $c$, complexity field, using light-cones of depth 3.  In $b$ and $c$,
    higher values of the field are denoted by darker points.  In $c$, the
    vertical grey band on the left is due to the fact that we need a past to
    compute the causal states.  There are a lot of particles in the beginning,
    which makes the background domain rather unusual and complex, but it
    gradually bleaches.}
  \label{fig:rule110_results}
\end{figure}

\begin{figure}[tbhp]
  \centering
$a$ \includegraphics[width=0.95\columnwidth]{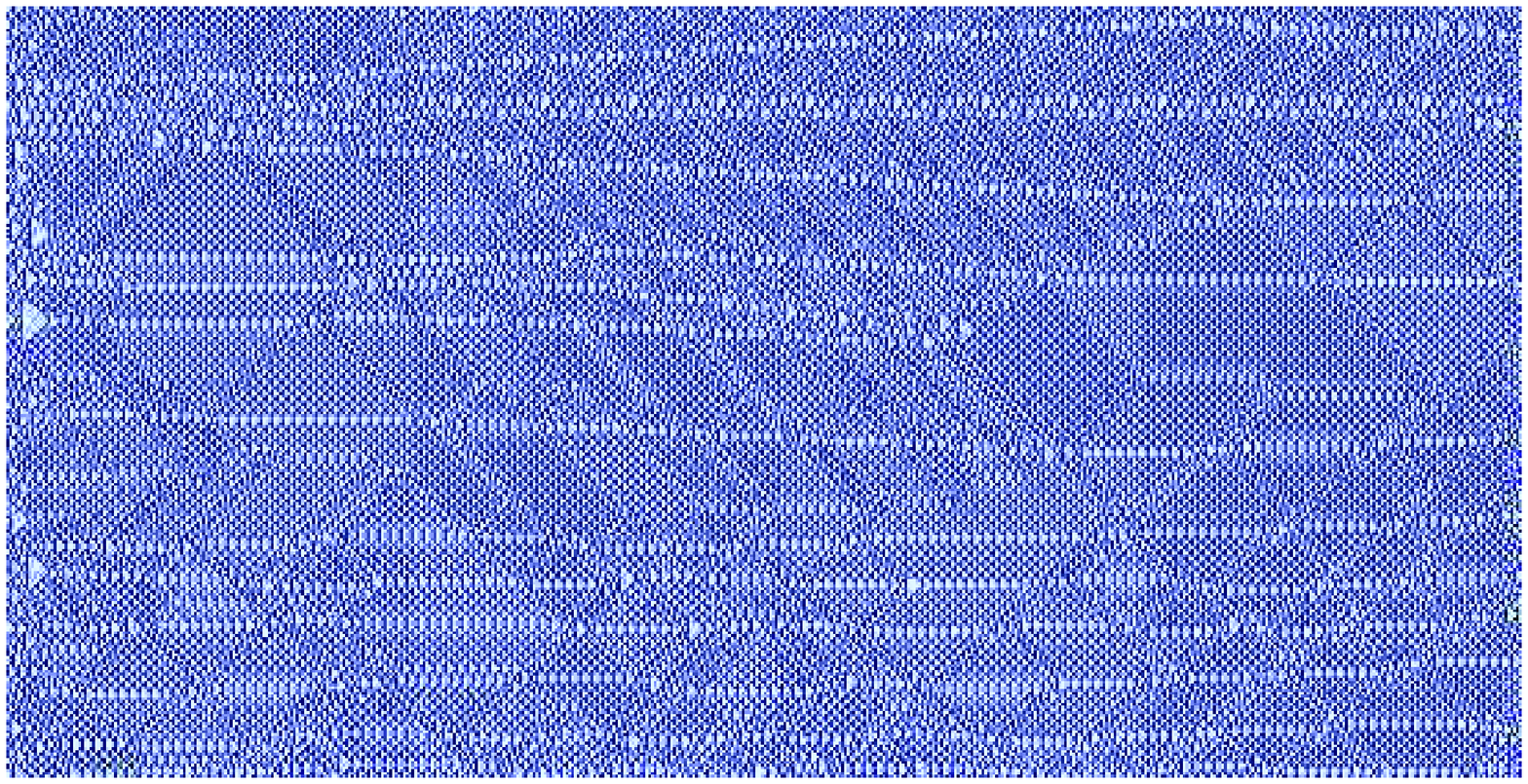}  
  \par\smallskip
$b$ \includegraphics[width=0.95\columnwidth]{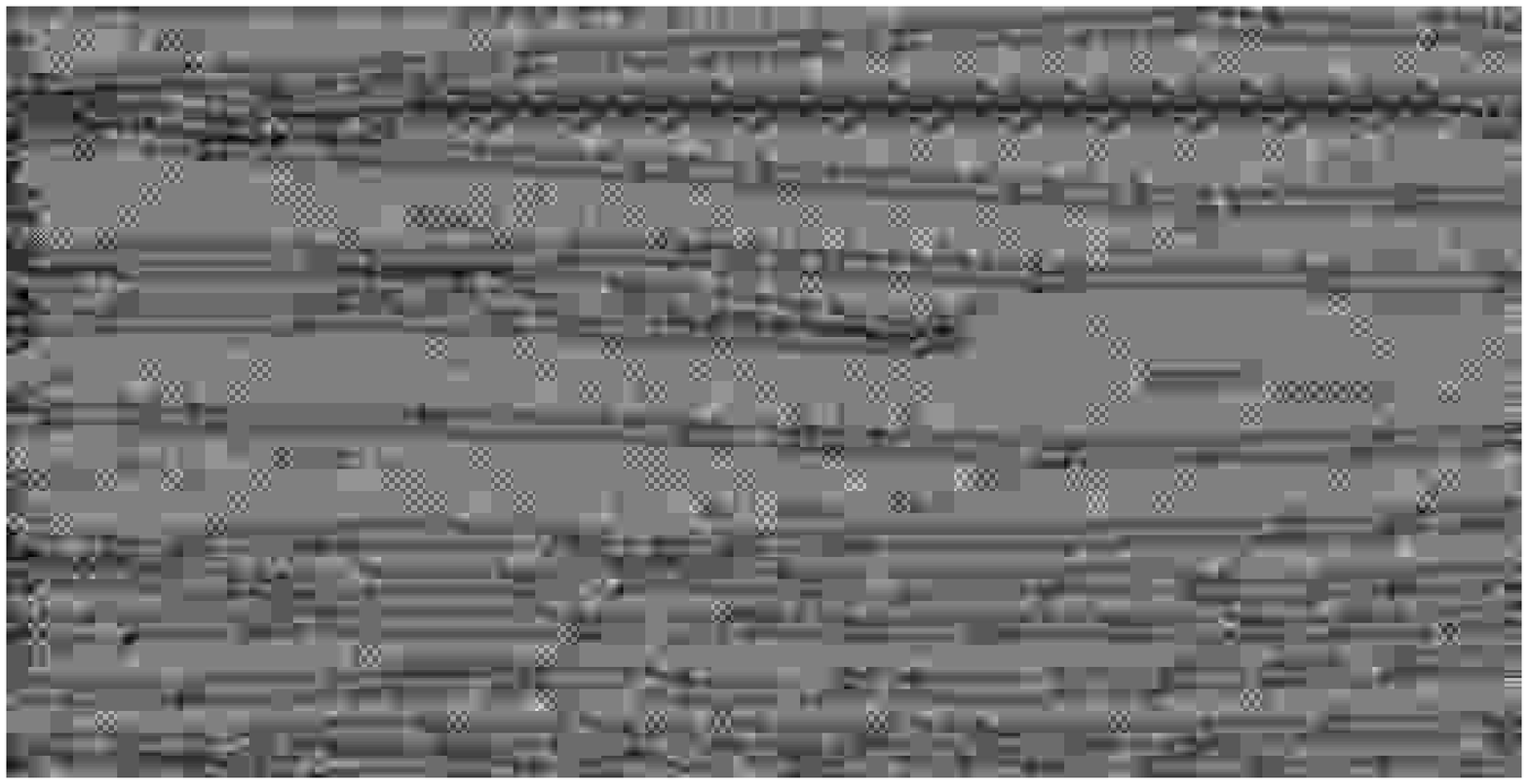}
  \par\smallskip
$c$ \includegraphics[width=0.95\columnwidth]{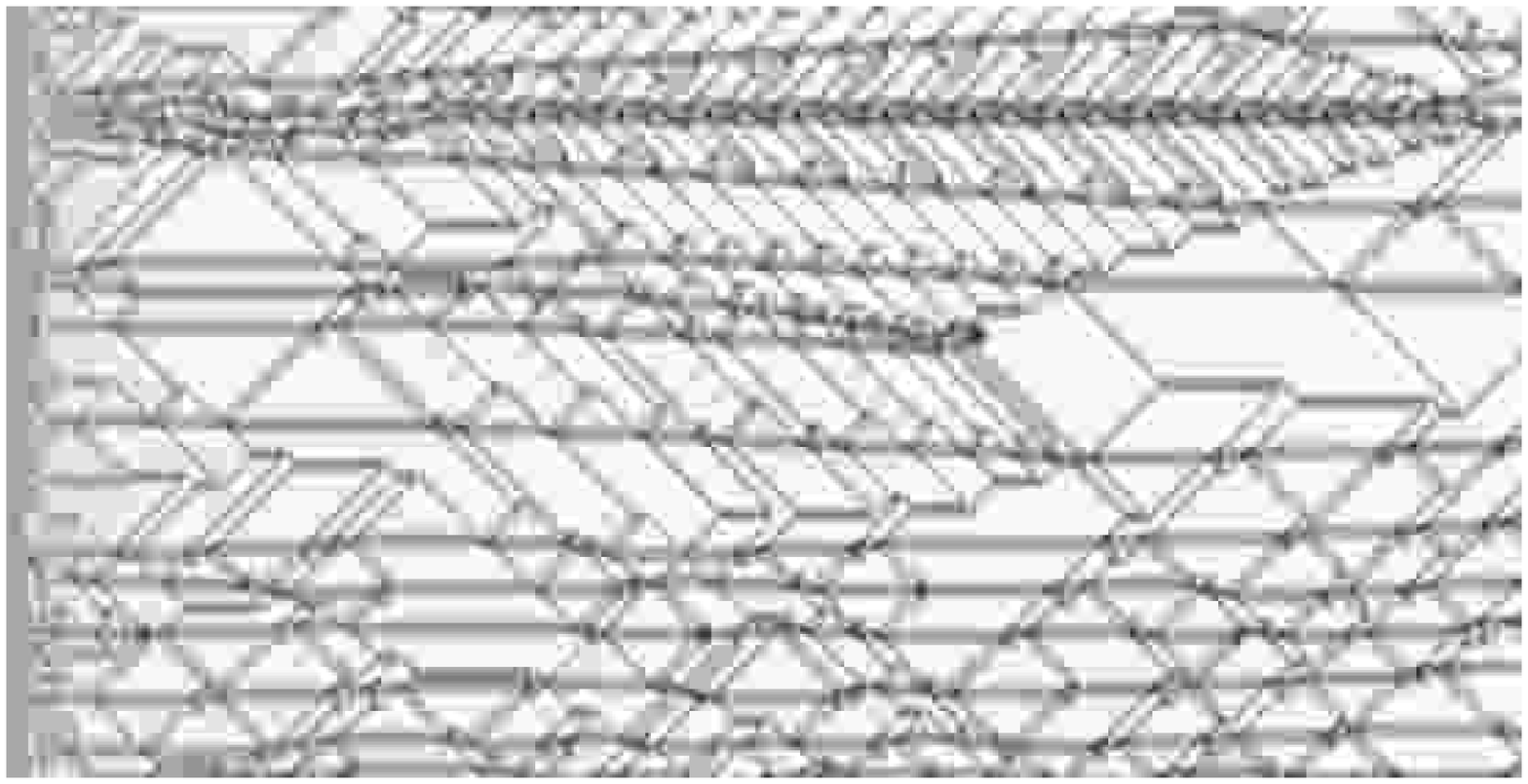}
  \caption{Evolution of rule 54 from a random initial condition: $a$,
    configuration field; $b$, local sensitivity ($p=1,\ f=10$); $c$,
    statistical complexity for rule 54 (light-cone depth of 3). In $c$, note
    that the background domain is light (low complexity), the particle are
    grey, and most of the collisions are darker (highest complexity).  The tiny
    particles that are merely phase shifts in the periodic background are made
    clear here, while they are hard to identify by eye in the configuration
    field (though misalignment between that field and your printer matrix can
    help!).}
   \label{fig:rule54_results}
\end{figure}

We now present the results of applying the dynamical sensitivity filter to
several 1+1D elementary CA.  In Figure \ref{fig:rule110_results} we show the
configuration field of rule 110 ($a$) and the field filtered with respect to
the sensitivity ($b$); Figure \ref{fig:rule54_results} does the same for rule
54.  The filtering automatically distinguishes the autonomous features of the
CA, in particular the particles' evolution through time.  The autonomy of the
particles is reflected by their dark tone.  The domains, in contrast, are much
less autonomous, and appear lighter.  This makes intuitive sense.  Domains are
large, ordered regions of the CA, and their perturbation has little effect upon
the dynamics, because any defects are either quickly healed or remain confined.
Particles in contrast are relatively complex and localized objects which travel
through the domains.  Perturbing a particle generally strongly changes the
dynamics of the CA, either through its destruction, or via a significant
alteration of its attributes---location, internal phase, type of particle,
etc. \cite{Hordijks-rule}.

\begin{figure}[tbhp]
  \centering
  $a$ \includegraphics[width=0.95\columnwidth]{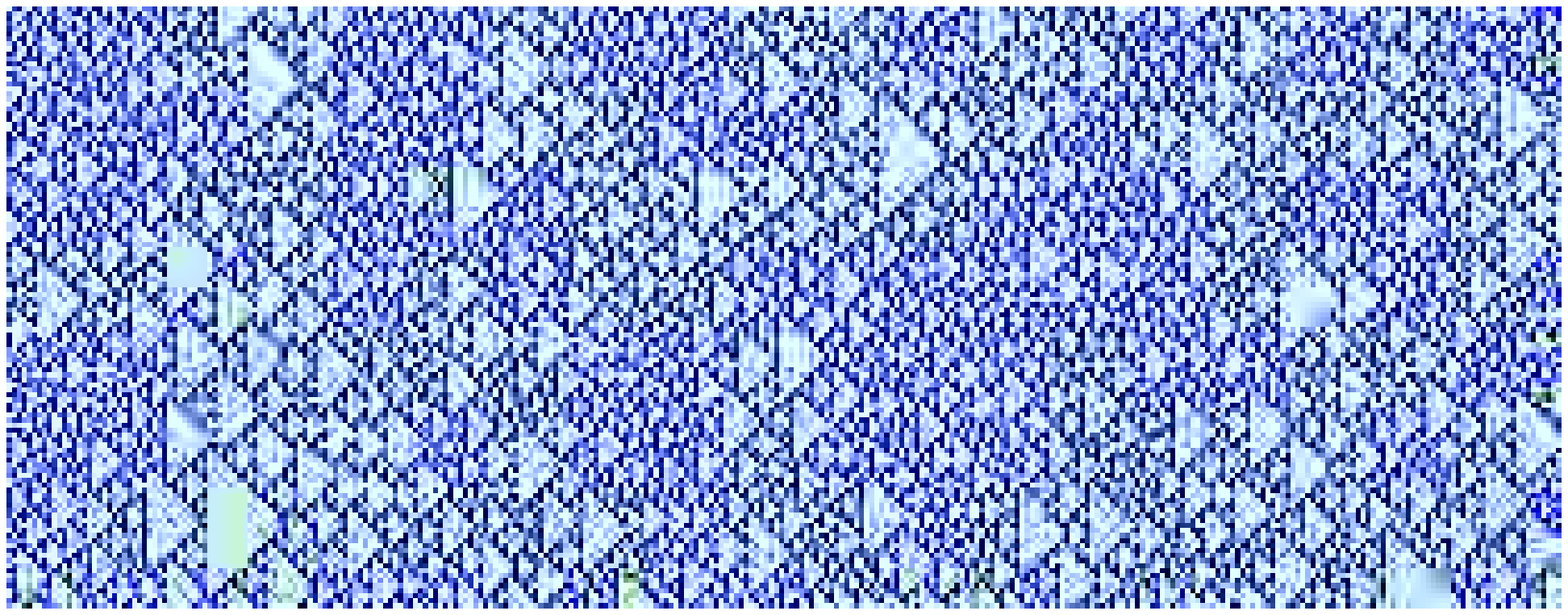}
  \par\smallskip
  $b$ \includegraphics[width=0.95\columnwidth]{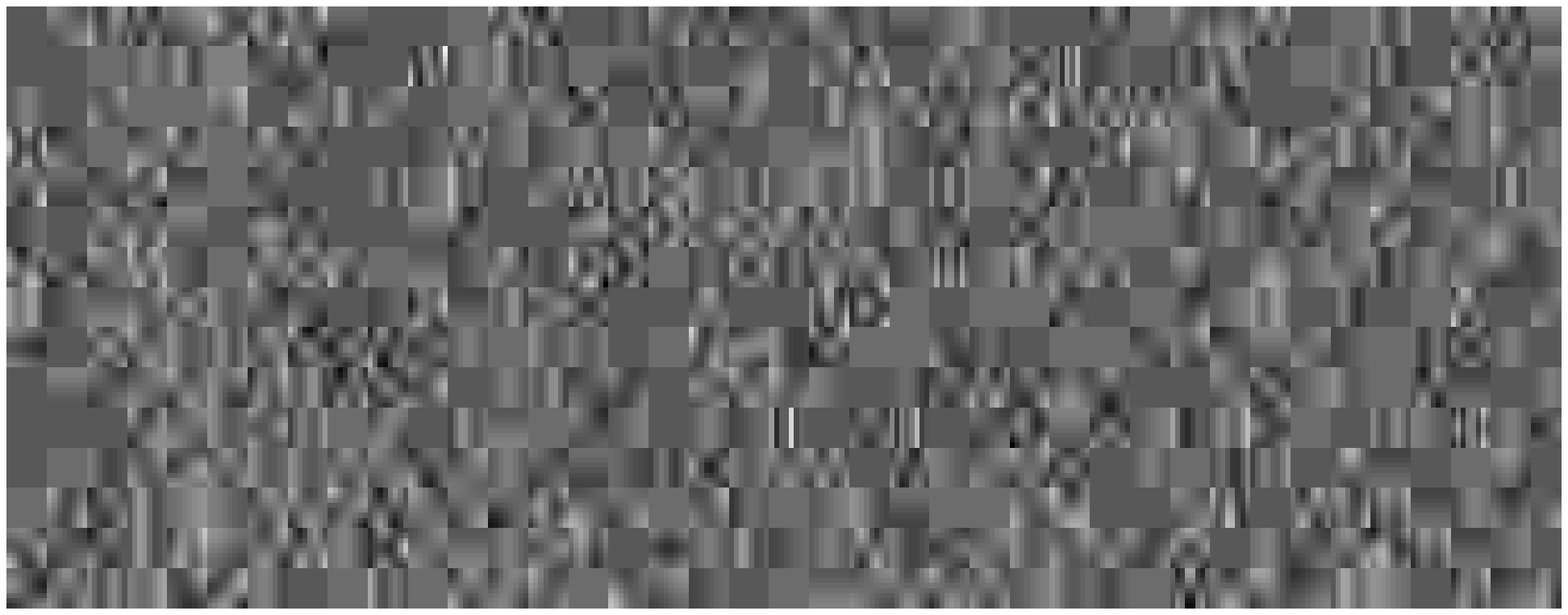}
  \par\smallskip
  $c$ \includegraphics[width=0.95\columnwidth]{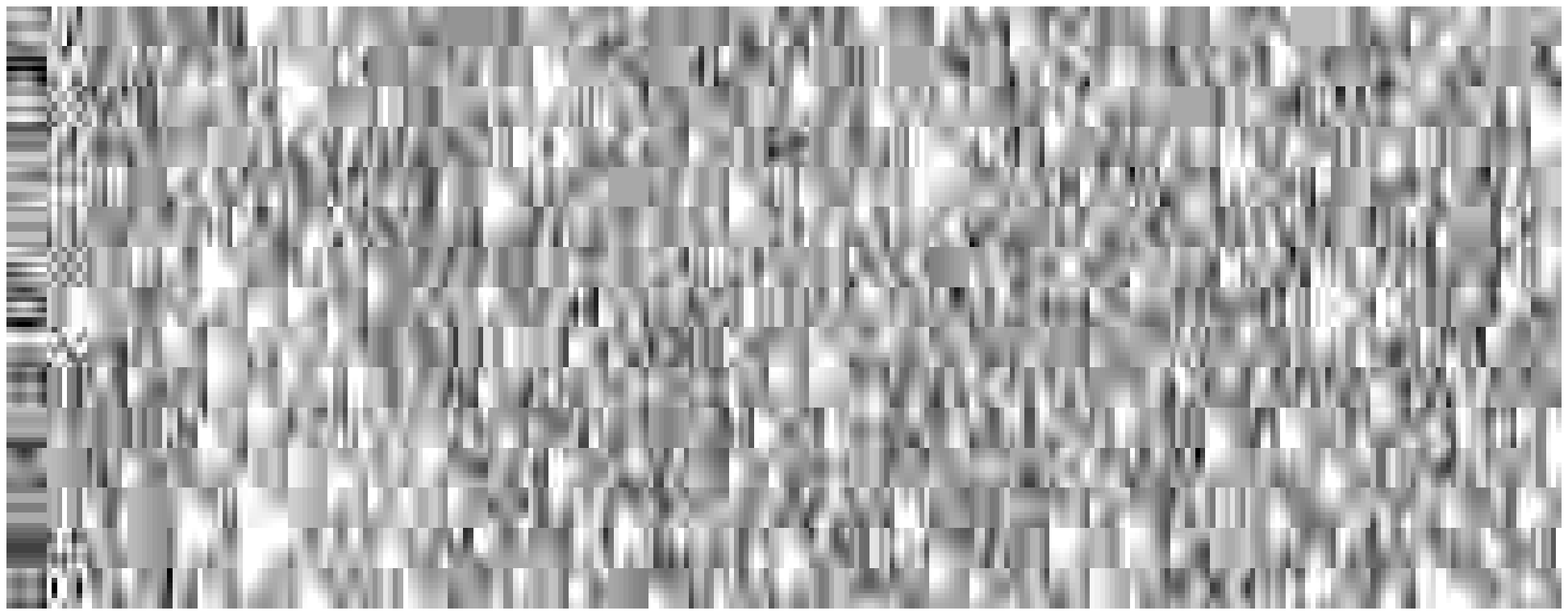}
  \caption{The evolution of rule 22 from a random initial condition: $a$,
    configuration field; $b$, local sensitivity (calculated with $p=1,\ f=10$);
    $c$, statistical complexity.  The nearly-uniform sensitivity field in $b$
    reflects the chaotic nature of the rule. }
  \label{fig:rule22_results}
\end{figure}

\begin{figure}[tbhp]
  \centering
  $a$ \includegraphics[width=0.95\columnwidth]{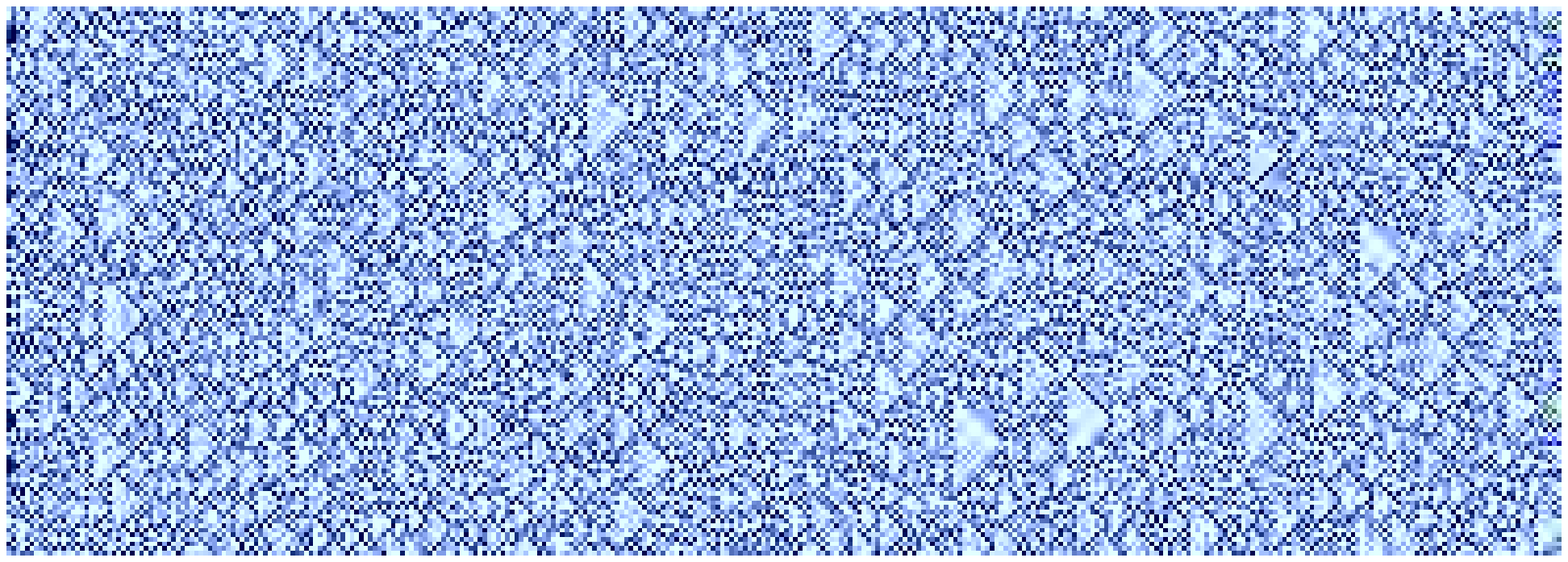}
  \par\smallskip
  $b$ \includegraphics[width=0.95\columnwidth]{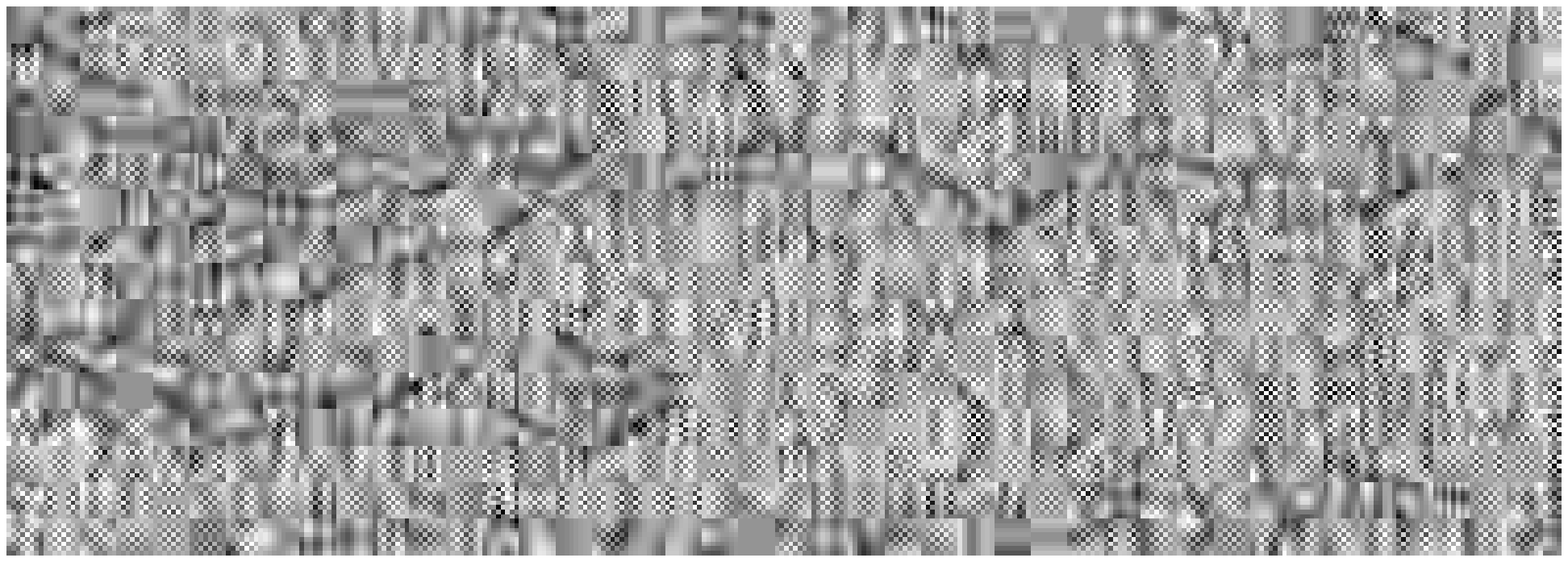}
  \par\smallskip
  $c$ \includegraphics[width=0.95\columnwidth]{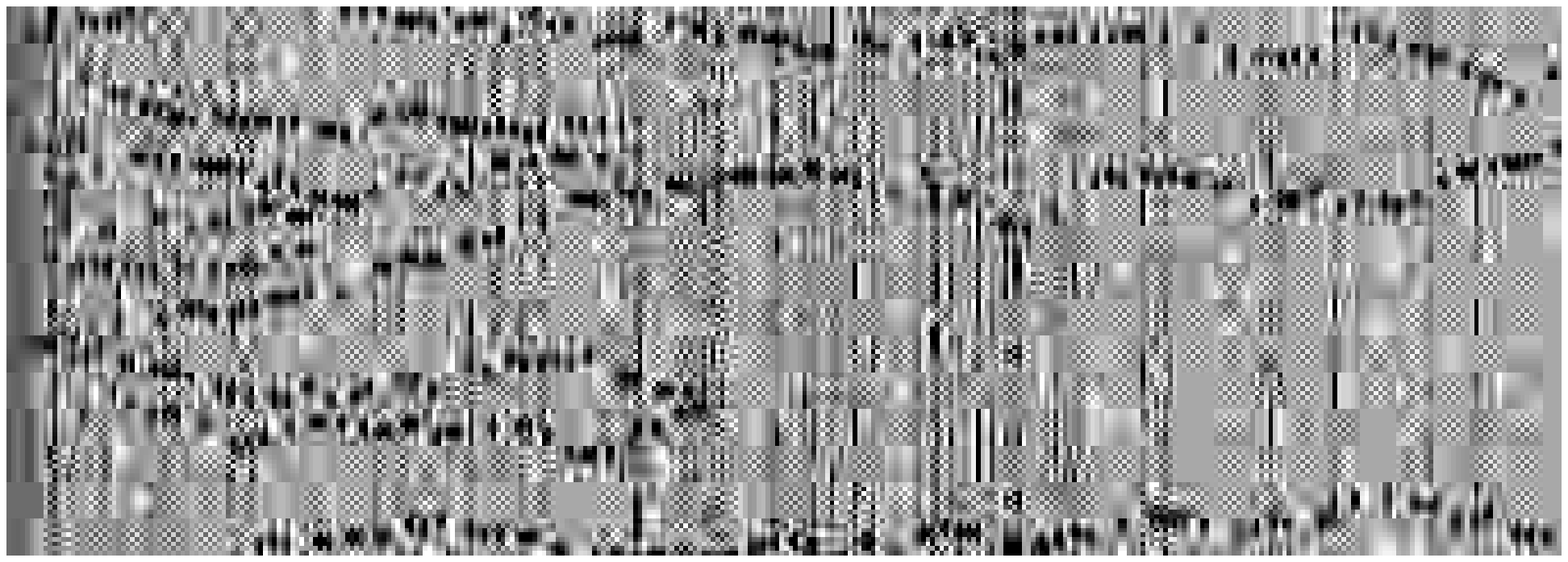}
  \caption{The evolution of rule 146 from a random initial condition: $a$, the
    configuration field; $b$, the sensitivity field (calculated using $p=1$ and
    $f=30$); $c$, the complexity field. Although the configuration field looks very
    similar to that of rule 22 (Figure \ref{fig:rule22_results}) this field is
    not chaotic and has a domain structure. Local sensitivity filtering $(b)$
    reveals the domain walls as light (low sensitivity) traces bounded by
    darker zones.  The domain walls aren't autonomous, their behavior is
    instead determined by what happens inside the domains, hence their low
    sensitivity.  Note also the increase in sensitivity when two walls are
    near: a small perturbation can lead them to merge (or prevent them from
    doing so). Statistical complexity filtering $(c)$ reveals the domain walls
    as dark (high complexity) traces composed of localized triangular
    regions. Because the domain walls require more predictive information than
    the domains themselves, the statistical complexity is higher for the walls.
    The dark vertical lines are finite size effects and should be ignored}
  \label{fig:rule146_results}
\end{figure}

The utility of the local sensitivity filter can be seen by comparing the raw
configurations produced by rules 22 and 146 to their filtered fields (Figures
\ref{fig:rule22_results} and \ref{fig:rule146_results}, respectively).  The
configuration fields of the two rules look very similar, both appearing highly
disordered.  Once filtered, however, it is clear that they are quite different.
Rule 22 is chaotic, and all points have roughly equal, and strong, influence on
the future of the system.  All of this, along with the absence of autonomous
objects, is plain from the sensitivity filter (Figure
\ref{fig:rule22_results}$b$).  In contrast, rule 146 (Figure
\ref{fig:rule146_results}) is composed of domains separated by boundaries which
wander over time.  The domain walls appear as light traces, indicating that
they are less autonomous than the domains they separate.  The domain walls of
rule 146 are dependent objects because their motion is determined by the
chaotic dynamics of the domains to either side of them.  This is a clear
instance of the fact that while the presence of an autonomous object implies
the existence of structures, the presence of structures does not necessarily
imply autonomy.

It is worth noting that in linear CA rules (those which are a sum (mod 2) on a
subset of the neighborhood) the local sensitivity is uniform over space.  This
is because the Hamming distance between the original and perturbed
configurations is independent of the perturbed cell, which in turn is because
the difference plume always has the same shape.  As expected from this
argument, direct calculations of the sensitivity field on additive rules such
as ECA 60, 85, 90, 105, 150 and 170 produce perfectly uniform results.  (We
omit plots for reasons of visual monotony.)

\section{Local Statistical Complexity}
\label{sec:sc}

Although the sensitivity is very effective at uncovering structures, as well as
telling us how stable those structures are, it has a significant drawback.  The
CA configuration field must be actively perturbed many times and the results
compared to the unperturbed case.  Such a procedure is computationally
intensive and often impossible to apply effectively to experimental data.  We
now discuss the {\it local statistical complexity} $C(\vec{r},t)$
\cite{CRS-thesis,CRS-prediction-on-networks,QSO-in-PRL}, the calculation of
which only requires observations, rather than active and repeated
perturbations.  We shall see that the statistical complexity has other
advantages as well.

\subsection{Local Causal States and Their Complexity}

Let $x(\vec{r},t)$ be an $n+1$D field, possibly stochastic, in which
interactions between different space-time points propagate at speed $c$.  As in
\cite{Parlitz-Merkwirth-local-states}, define the {\em past light cone} of the
space-time point $(\vec{r}, t)$ as all points which could influence
$x(\vec{r},t)$, i.e., all points $(\vec{q}, u)$ where $u < t$ and $||\vec{q} -
\vec{r}|| \leq c(t-u)$.  The {\em future light cone} of $(\vec{r},t)$ is the
set of all points which could be influenced by what happens at $(\vec{r},t)$.
$\localpast(\vec{r},t)$ is the configuration of the field in the past light
cone, and $\localfuture(\vec{r},t)$ the field in the future light cone.  The
distribution of future light cone configurations, given the configuration in
the past, is $\Prob(\localfuture|\localpast)$.

Any function $\eta$ of $\localpast$ defines a {\em local statistic}.  It
summarizes the influence of all the space-time points which could affect what
happens at $(\vec{r},t)$.  Such local statistics should tell us something about
``what comes next,'' which is $\localfuture$.  Information theory lets us
quantify how informative different statistics are, and so guides our choice
among them.

The information about variable $x$ in variable $y$ is
\begin{eqnarray}
I[x;y] & \equiv & \left\langle \log_2{\frac{\Prob(x,y)}{\Prob(x)\Prob(y)}}
\right\rangle
\end{eqnarray}
where $\Prob(x,y)$ is joint probability, $\Prob(x)$ is marginal probability,
and $\langle \cdot \rangle$ is expectation
\cite{Kullback-info-theory-and-stats}.  The information a statistic $\eta$
conveys about the future is $I[\localfuture;\eta(\localpast)]$.  A statistic is
{\em sufficient} if it is as informative as possible
\cite{Kullback-info-theory-and-stats}, here if and only if
$I[\localfuture;\eta(\localpast)] = I[\localfuture;\localpast]$.  This is the
same \cite{Kullback-info-theory-and-stats} as requiring that
$\Prob(\localfuture|\eta(\localpast)) = \Prob(\localfuture|\localpast)$.  A
sufficient statistic retains all the predictive information in the data.  Since
we want {\em optimal} prediction,we confine ourselves to sufficient statistics.

If we use a sufficient statistic $\eta$ for prediction, we must describe or
encode it.  Since $\eta(\localpast)$ is a function of $\localpast$, this
encoding takes $I[\eta(\localpast);\localpast]$ bits.  If knowing $\eta_1$ lets
us compute $\eta_2$, which is also sufficient, then $\eta_2$ is a more concise
summary, and $I[\eta_1(\localpast);\localpast] \geq
I[\eta_2(\localpast);\localpast]$.  A {\em minimal sufficient statistic}
\cite{Kullback-info-theory-and-stats} can be computed from any other sufficient
statistic.  In the present context, the minimal sufficient statistic is
essentially unique (as discussed below), and can be reached through the
following construction.

Take two past light cone configurations, $\localpast_1$ and $\localpast_2$.
Each has some conditional distribution over future light cone configurations,
$\Prob(\localfuture|\localpast_1)$ and $\Prob(\localfuture|\localpast_2)$
respectively.  The two past configurations are equivalent, $\localpast_1 \sim
\localpast_2$, if those conditional distributions are equal.  The set of
configurations equivalent to $\localpast$ is $[\localpast]$.  Our statistic is
the function which maps past configurations to their equivalence classes:
\begin{equation}
\epsilon(\localpast)  \equiv [\localpast]
 =  \left\{\localpastprime ~: ~\Prob(\localfuture|\localpastprime) = \Prob(\localfuture|\localpast)\right\}
\end{equation}
Clearly, $\Prob(\localfuture|\epsilon(\localpast)) =
\Prob(\localfuture|\localpast)$, and so $I[\localfuture;\epsilon(\localpast)] =
I[\localfuture;\localpast]$, making $\epsilon$ a sufficient statistic.  The
equivalence classes, the values $\epsilon$ can take, are the {\em causal
  states} \cite{CRS-prediction-on-networks,Inferring-stat-compl,CMPPSS,%
  What-is-a-macrostate}.  Each causal state is a set of specific past
light-cones, and all the cones it contains are equivalent, predicting the same
possible futures with the same probabilities.  Thus there is no advantage to
subdividing the causal states, which are the coarsest set of predictively
sufficient states.

For any sufficient statistic $\eta$, $\Prob(\localfuture|\localpast) =
\Prob(\localfuture|\eta(\localpast))$.  So if $\eta(\localpast_1) =
\eta(\localpast_2)$, then $\Prob(\localfuture|\localpast_1) =
\Prob(\localfuture|\localpast_2)$, and the two pasts belong to the same causal
state.  Since we can get the causal state from $\eta(\localpast)$, we can use
the latter to compute $\epsilon(\localpast)$.  Thus, $\epsilon$ is minimal.
Moreover, $\epsilon$ is the {\em unique} minimal sufficient statistic
\cite[Corollary 3]{CRS-prediction-on-networks}: any other just relabels the
same states.

Because $\epsilon$ is minimal, $I[\epsilon(\localpast);\localpast] \leq
I[\eta(\localpast);\localpast]$, for any other sufficient statistic $\eta$.
Thus we can speak objectively about the minimal amount of information needed to
predict the system, which is how much information about the past of the system
is relevant to predicting its own dynamics.  This quantity,
$I[\epsilon(\localpast);\localpast]$, is a characteristic of the system, and
not of any particular model.  We define the {\em local statistical complexity} as
\begin{eqnarray}
C(\vec{r},t) & \equiv & I[\epsilon(\localpast(\vec{r},t));\localpast(\vec{r},t)]
\end{eqnarray}
For a discrete field, $C$ is also equal to $H[\epsilon(\localpast)]$, the
Shannon entropy of the local causal state\footnote{The proof is as
  follows. $I[\epsilon(\localpast);\localpast] = H[\epsilon(\localpast)] -
  H[\epsilon(\localpast)|\localpast]$, the amount by which the uncertainty in
  $\epsilon(\localpast)$ is reduced by knowing $\localpast$.  But for any
  discrete-valued function $f$, $H[f(x)|x] = 0$, because a function is certain,
  given its argument.  Hence $I[\epsilon(\localpast);\localpast] =
  H[\epsilon(\localpast)]$.}.  $C$ is the amount of information required to
describe the behavior at that point, and equals the log of the effective number
of causal states, i.e., of different distributions for the future.  Complexity
lies between disorder and order
\cite{Badii-Politi,Grassberger-1986,Inferring-stat-compl}, and $C = 0$ both
when the field is completely disordered (all values of $x$ are independent) and
completely ordered ($x$ is constant). $C$ grows when the field's dynamics
become more flexible and intricate, and more information is needed to describe
the behavior.

The local complexity field $C(\vec{r},t)$, is simply $-\log{\Pr(s(\vec{r},t))}$
where $s(\vec{r},t)$ is the local causal state at space-time point
$(\vec{r},t)$.  The local complexity is the number of bits which would be
required to encode the causal state at $\vec{r}, t$, if we used the optimal
(Shannon) coding scheme.  Equivalently, it is the number of bits of information
about $\localpast(\vec{r},t)$ which are used to determine the causal state.

It is appropriate, at this point, to take a step back and consider what we are
doing.  Why should we use the light-cone construction, as opposed to any other
kind of localized predictor?  Indeed, why use localized statistics at all,
rather than global methods?  Let us answer these in reverse order.  The use of
local predictors is partly a matter of interest---in studying coherent
structures, we care essentially about spatial organization, and so global
approaches, which would treat the system's sequence of configurations as one
giant time series, simply don't tell us what we want to know.  In part, too,
the local approach makes a virtue of necessity, because global prediction
quickly becomes impractical for systems of any real size.  The number of modes
required by methods attempting global prediction, like Karhunen-Loeve
decomposition, grows extensively with system volume
\cite{Zoldi-Greenside-extensive-chaos,Zoldi-et-al-extensive-scaling}.  Global
methods thus provide no {\em advantage} in terms of compression or accuracy.

The use of light-cones for the local predictors, first suggested by
\cite{Parlitz-Merkwirth-local-states},\footnote{It seems that the use of light
  cones to analyze spatial stochastic processes was first suggested by
  Kolmogorov, who called them ``causal sets'', in the context of a model of
  crystallization \cite{Kolmogorov-light-cones}---see
  \cite{Capasso-Micheletti-spatial-birth-growth}.} rather than some other
shape, is motivated partly by physical considerations, and partly the nice
formal features which follow from the shape, of which we will mention three
\cite{CRS-prediction-on-networks}.
\begin{enumerate}
\item The light-cone causal states, while local statistics, do not lose any
  global predictive power.  To be precise, if we specify the causal state at
  each point in a spatial region, that array of states is itself a sufficient
  statistic for the future configuration of the region, even if the region is
  the entire lattice.
\item The light-cone states can be found by a recursive filter.  To illustrate
  what this means, consider two space-time points, $(\vec{r},t)$ and
  $(\vec{q},u)$  $u \geq t$.  The state at each point is determined by the
  configuration in its past light-cone: $\localstate(\vec{r},t) =
  \epsilon(\localpast(\vec{r},t))$, $\localstate(\vec{q},u) =
  \epsilon(\localpast(\vec{q},u))$.  The recursive-filtration property means
  that we can construct a function which will give us $\localstate(\vec{q},u)$
  as a function of $\localstate(\vec{r},t)$, plus the part of the past
  light-cone of $(\vec{q},u)$ that is not visible from $(\vec{r},t)$.  Not only
  does this greatly simplify state estimation, it opens up powerful connections
  to the theory of two-dimensional automata \cite{Two-D-Patterns}.
\item The local causal states form a Markov random field, once again allowing
  very powerful analytical techniques to be employed which would not otherwise
  be available \cite{Kindermann-Snell-on-Markov-random-fields}.
\end{enumerate}
In general, if we used some other shape than the light-cones, we would not
retain any of these properties.

\subsection{A Reconstruction Algorithm}

We now sketch an algorithm to recover the causal states from data, and so
estimate $C$.  (See Fig.\ \ref{algorithm} for pseudo-code,
Ref.\ \cite{CRS-prediction-on-networks} for details, and
cf.\ \cite{Parlitz-Merkwirth-local-states}.)  At each time $t$, list the
observed past and future light-cone configurations, and put the observed past
configurations in some arbitrary order, $\left\{\localpast_i\right\}$.  (In
practice, we must limit how far light-cones extend into the past or future.)
For each past configuration $\localpast_i$, estimate
$\Prob_t(\localfuture|\localpast_i)$.  We want to estimate the states, which
ideally are groups of past cones with the same conditional distribution over
future cone configurations.  Not knowing the conditional distributions a
priori, we must estimate them from data, and with finitely many samples, such
estimates always have some error.  Thus, we approximate the true causal states
by clusters of past light-cones with similar distributions over future
light-cones; the conditional distribution for a cluster is the weighted mean of
those of its constituent past cones.  Start by assigning the first past,
$\localpast_1$ to the first cluster.  Thereafter, for each $\localpast_i$, go
down the list of existing clusters and check whether
$\Prob_t(\localfuture|\localpast_i)$ differs significantly from each cluster's
distribution, as determined by a fixed-size $\chi^2$ test.  (We used $\alpha =
0.05$ in all our calculations.)  If the discrepancy is insignificant, add
$\localpast_i$ to the first matching cluster, updating the latter's
distribution.  Make a new cluster if $\localpast_i$ does not match any existing
cluster.  Continue until every $\localpast_i$ is assigned to some cluster.  The
clusters are then the estimated causal states at time $t$.  Finally, obtain the
probabilities of the different causal states from the empirical probabilities
of their constituent past configurations, and calculate $C(t)$.  This procedure
converges on the correct causal states as it gets more data, independent of the
order of presentation of the past light-cones, the ordering of the clusters, or
the size $\alpha$ of the significance test \cite{CRS-prediction-on-networks}.
For finite data, the order of presentation matters, but we finesse this by
randomizing the order.

\begin{figure}[t]
\begin{tabbing}
\texttt{U} $\leftarrow$ list of all pasts in random order\\
Move the first past in \texttt{U} to a new state\\
for \= each \texttt{past} in \texttt{U}\\
\> \texttt{noMatch} $\leftarrow$ TRUE\\
\> \texttt{state} $\leftarrow$ first state on the list of states\\
\> while \= (\texttt{noMatch} and more states to check)\\
\> \> \texttt{noMatch} $\leftarrow$ (\=Significant difference between \\
\> \> \> \texttt{past} and \texttt{state}?)\\
\> \> if \= (\texttt{noMatch})\\
\> \> \> \texttt{state} $\leftarrow$ next state on the list\\
\> \> else \=\\
\> \> \> Move \texttt{past} from \texttt{U} to \texttt{state}\\
\> \> \> \texttt{noMatch} $\leftarrow$ FALSE\\
\> if \= (\texttt{noMatch})\\
\> \> make a new state and move \texttt{past} into it from \texttt{U}
\end{tabbing}
\caption{\label{algorithm} Algorithm for grouping past light-cones into estimated states}
\end{figure}

\subsection{Results on Elementary Cellular Automata}
\label{sec:sc-eca}

In \S \ref{sec:lle-eca}, we demonstrated the ability of local sensitivity
filtering to find coherent structures in elementary cellular automata.  Here we
apply local statistical complexity filtering to the same cellular automata, and
compare the results of the two procedures.

Figures \ref{fig:rule110_results}$c$ and \ref{fig:rule54_results}$c$ show the
local complexity fields $C(\vec{r},t)$ of two of the classic rules, 110 and 54,
respectively.  Particles stand out clearly and distinctly on clean backgrounds.
Note that we achieve this result by applying the same filter to both systems,
even though their background domains and particles are completely different.
Were one to use, say, the conventional regular-language filter constructed in
\cite{Comp-mech-of-CA-example} for rule 54 on rule 110, it would produce
nonsense.  (Ref.\ \cite{Hordijks-rule} discusses the domains, particles and
conventional filters of both these rules.)  Note that the filter in
\cite{Comp-mech-of-CA-example} is hand-crafted, based on a detailed
understanding of the CA's dynamics, whereas our filter, as we have said, is
completely automatic and requires no human intervention.  One might expect this
generality to be paid for in a loss of resolving power, or missing
system-specific features, but this does not appear to be the case\footnote{This
  statement must be qualified by a recognition that we must supply the filter
  with enough data for it to find the right states.  The learning rate of the
  state reconstruction algorithm is an important topic, beyond the scope of
  this paper.}.  For instance, the regular-language analysis of rule 54
\cite{Comp-mech-of-CA-example} identifies a subtle kind of particle, which
consists of a phase shift in the spatially-periodic background domain.  These
phase shifts are hard to identify by eye, but show up very cleanly as particles
in Figure \ref{fig:rule54_results}$c$.

For completeness, and further comparison with the local sensitivity, we include
rules 22 and 146 filtered with respect with statistical complexity in Figures
\ref{fig:rule22_results}$c$ and \ref{fig:rule146_results}$c$.  Statistical
complexity is also able to distinguish between the two CA configuration fields, even
though the difference is not readily apparent to the naked eye.

Experimentally, we find that the depth of the past light cone is more relevant
to proper filtering than the depth of the future light cone.  (This may be
related to the recursive estimation property of the causal states.)  On
elementary cellular automata, depth 2 is often sufficient, in the sense that
further extensions of the cones do not change the states identified, but the
results presented here use depth 3 for both future and past light cones.  (Rule
41, not shown for reasons of space, required a past-depth of 5 in order to
reach convergence.)

While both our techniques make it easy to identify objects like particles, even
when they were previously hard to detect, they are very different filters, not
just in their definition but also in their results.  This can immediately be
seen from the correlation coefficients\footnote{The correlation coefficient of
  statistics, $\rho_{xy} = \frac{\langle XY \rangle - \langle X \rangle \langle
    Y \rangle}{\sigma_{X} \sigma_{Y}}$, is a dimensionless counterpart to the
  correlation function of statistical physics, $C_{XY} = \langle XY \rangle -
  \langle X \rangle \langle Y \rangle$, normalized so that its value lies
  between $-1$ and $+1$, and is zero when $X$ and $Y$ are linearly unrelated.}
of the sensitivity and complexity fields---for rule 110, for instance, the
correlation is a negligible $0.014$.

More abstractly, statistical complexity is a local quantity that is calculated
using a global object, namely the probability distribution over causal states.
Its accurate estimation thus needs a quite large number of cells, since,
lacking an analytical form for that distribution, the latter must itself be
estimated.  A further consequence is that the separation between particles and
other structures and the background tends to become cleaner over time, as the
domains grow and the density of particles decays, suppressing the complexity of
the former and raising that of the later.  (This may be clearly seen in Figure
\ref{fig:rule110_results}$c$.)  Identification of structures from the
complexity field is thus best undertaken after a (hopefully short) transient
regime has passed, during which the local sensitivity filter may be more
useful.  It is not, of course, necessary for the system to have reached a
stationary regime in order to use the local complexity filter.

\section{Results on Cyclic Cellular Automata}
\label{sec:cca}

In this section we compare and contrast local sensitivity and statistical
complexity with a more traditional order parameter approach in a spiral-forming
CA model of an excitable medium.  By applying all three analyses to the same
2+1D CA, we will clarify the specific structural details emphasized by each
method, and bring out the advantages of automatic filtering over approaches
where the details must be put in by hand.

Our model system in this section is the cyclic cellular automata (CCA) on a
square lattice, which have been studied in considerable detail in the
literature on spatial stochastic processes
\cite{Fisch-Gravner-Griffeath-threshold-range,Fisch-Gravner-Griffeath}.  In the
general case, there are $\kappa$ colors, numbered from $0$ to $\kappa-1$.  A
cell of color $k$ changes its color only if at least $T$ of its neighbors are
of color $k+1 \bmod \kappa$, in which case it assumes that color.  In this
paper, we confine ourselves to the special case of range-one neighborhoods,
$\kappa = 4$, and $T=2$.  The behavior of the system, started from uniform
random initial conditions, is illustrated in Figure
\ref{fig:CCA-time-evolution}.  In \cite{QSO-in-PRL}, we demonstrated the
ability of statistical complexity to quantify the extent to which the CA
self-organizes, and the effect of changing parameter values on the degree of
self-organization.  Here, however, we are more interested in the patterns
formed than in whether significant pattern formation is taking place, so we
deal only with the most strongly self-organizing variant.

\begin{figure}[thbp]
\includegraphics[width=\columnwidth]{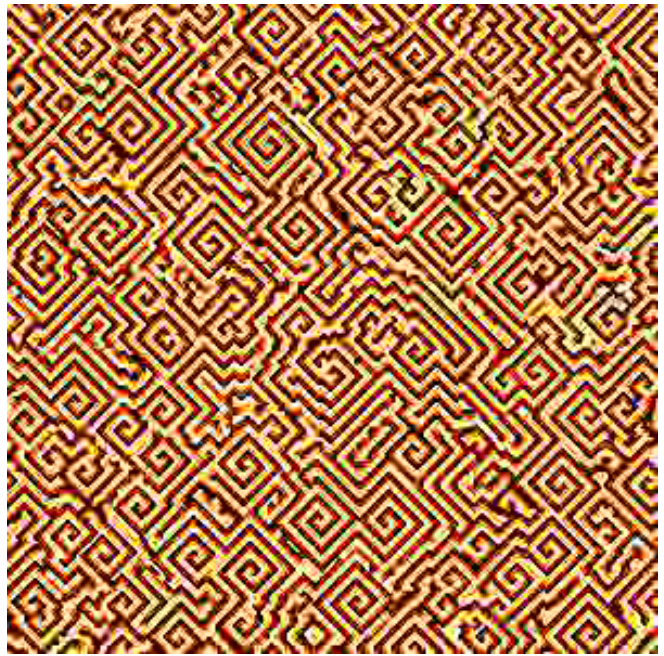}
\caption{(Color online.)  A typical configuration of the spiral-forming CCA
  cyclic cellular automaton in the asymptotic regime ($t=200$), illustrating
  the presence of coherent structures (rotating spiral waves) dominating the
  lattice.}
\label{fig:CCA-screenshot}
\end{figure}

\subsection{Order Parameter and Effective Free Energy}
\label{sec:cca-op}

While, as remarked, cyclic CA have been extensively studied as models of
excitable media, they have not (to the best of our knowledge) hitherto been
examined with the tools of the order parameter approach.  To better highlight
the distinctive characteristics of the automatic filtering methods we propose,
we first identify an order parameter and effective free energy for the present
version of CCA.  Our free energy will be a functional of the configuration field
configuration alone, and is most appropriate for the long-run stationary
distribution, rather than the initial transient stages, when the system is far
from statistical equilibrium.  Accordingly, we suppress time as an argument to
the configuration field in what follows.

\begin{figure}[thbp]
\centering \includegraphics[width=\columnwidth]{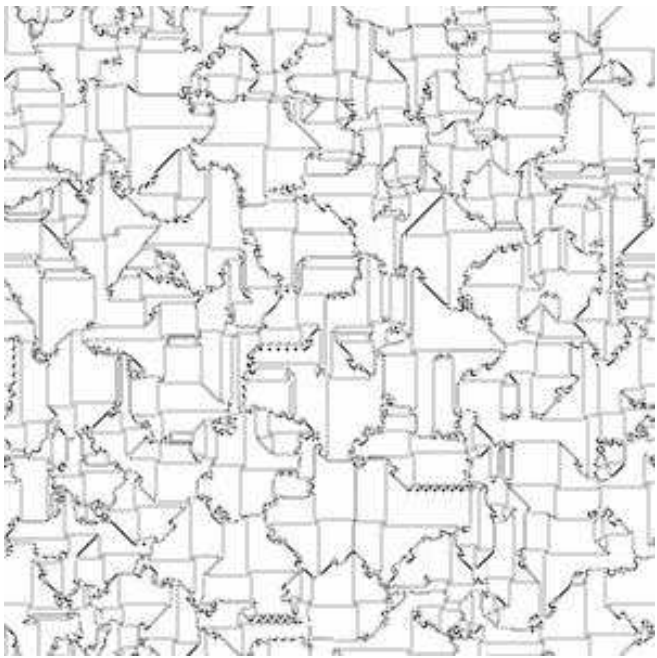}
\caption{Free energy per site of the CCA configuration field given in Figure
  \ref{fig:CCA-screenshot}.  Darker cells have higher energy values.  Note the
  increased free energy at the topological defects, e.g. the spiral cores and
  domain walls.}
\label{fig:free_energy}
\end{figure}

As noted, with these parameter settings, the CA configuration field forms rotating
spiral waves, which grow to engulf the entire lattice, with disordered domain
walls at the boundaries between competing spiral cores.  The mechanisms by
which spirals form and grow are fairly well understood, and topological
arguments \cite{Fisch-Gravner-Griffeath} pick out the key role of both
conservation of winding number and of the spiral cores in this process.
Accordingly we consider CCA as a kind of discretized $XY$ model, similar to a
clock model \cite[\S 3.6.3]{Chaikin-Lubensky}\footnote{A class of cellular
  automata very similar to CCA are treated as antiferromagnetic Potts models in
  \cite{Szolnoki-Szabo-Ravasz-on-Potts-model-rock-paper-scissors}.  We
  experimented with such an order parameter, but the results were poor.}, and,
as in such models generally, the appropriate order parameter has two components
\cite{Chaikin-Lubensky}.  Here the crucial observation is that the ground state
consists of plane waves, with stripes of cells of constant color extending
perpendicular to the direction of propagation of the wave.  The order parameter
is the local normalized wave-vector.
\begin{eqnarray*}
{\bf \Psi}(x,y) = \bar\Psi_x(x,y) \hat x + \bar\Psi_y(x,y) \hat y 
\end{eqnarray*}
defined such that it takes a one of four values ${\bf
  \Psi}=\frac{1}{\sqrt{2}}(\pm 1,\pm1)$ in each of the 4 domains surrounding a
spiral core.  The exact definition of the wave vector in terms of the states of
the CCA configuration field is given in appendix \ref{app:order-parameter}.  This wave
vector can be used to define a phase for the spiral wave at each lattice site.
\begin{eqnarray*}
\Phi(x,y)=\tan^{-1}\frac{\bar\Psi_y(x,y)}{\bar\Psi_x(x,y)}
\end{eqnarray*}
The local free energy is then calculated as the discretized version of
\begin{eqnarray}
F(x,y)={(\nabla\Phi(x,y))}^2
\end{eqnarray}
We show the free energy of the CA configuration field in Figure \ref{fig:free_energy}.
Note the increased free energy caused by topological defects such as the domain
walls and spiral cores.

\subsection{Local Sensitivity and Statistical Complexity for CCA}

Filtering by sensitivity and complexity, which we demonstrated above for
elementary cellular automata, works without modification on cyclic CA.  As
might be expected, the two filters reveal different, but compatible, aspects of
the system.

\begin{figure}[thbp]
  \centering
  \includegraphics[width=\columnwidth]{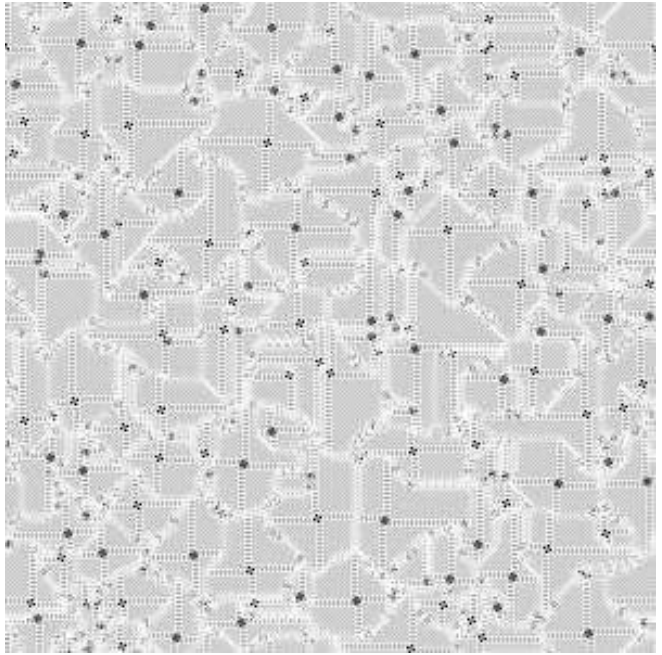}
  \caption{Local sensitivity ($p=0$, $f=5$) for the cyclic cellular automaton,
    using the CCA configuration field of Figure \ref{fig:CCA-screenshot}}
  \label{fig:cyclic-lle-lpr0}
\end{figure}

\begin{figure}[thbp]
  \centering
  \hfill
  $a$ \includegraphics[width=0.4\columnwidth]{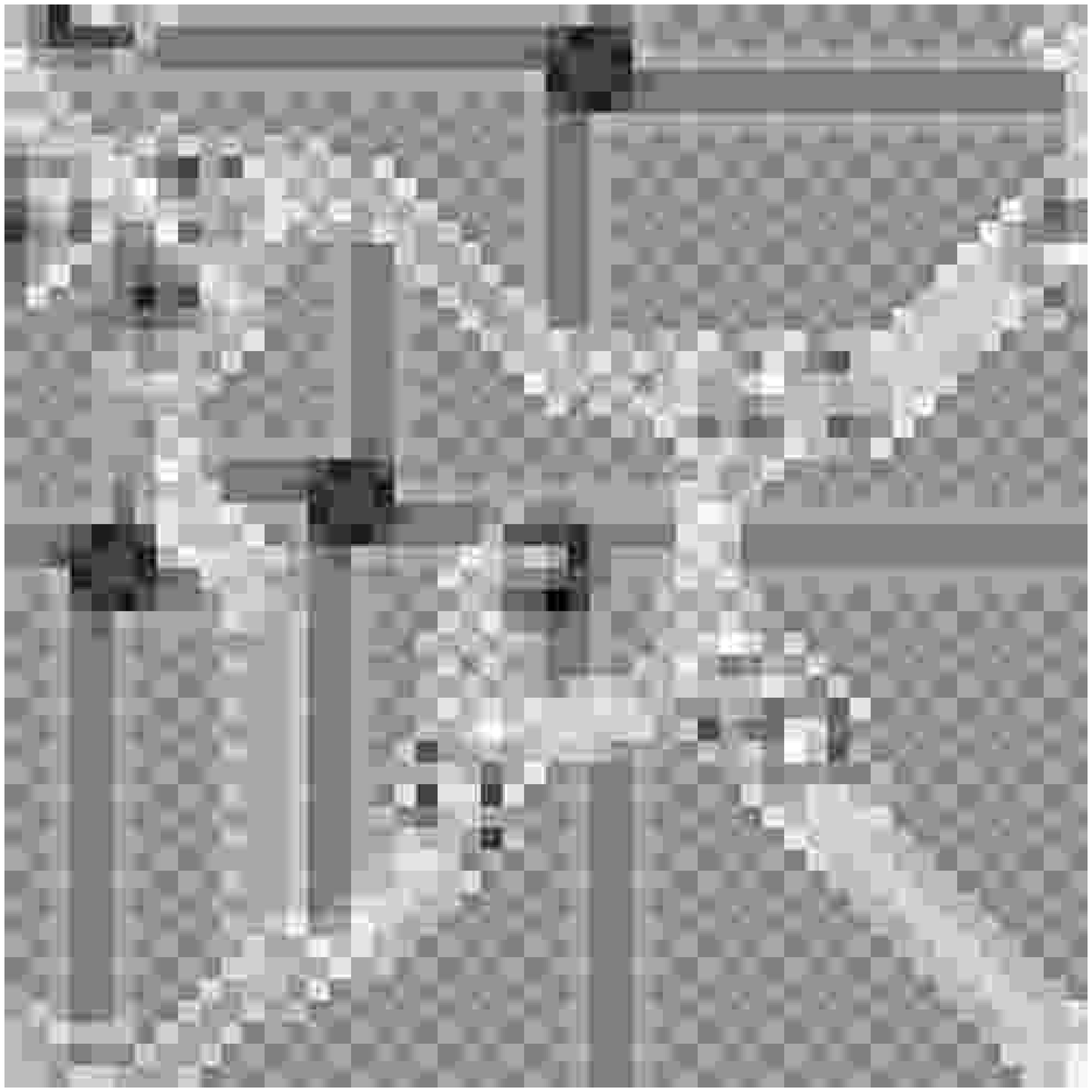}
  \hfill
  $b$ \includegraphics[width=0.4\columnwidth]{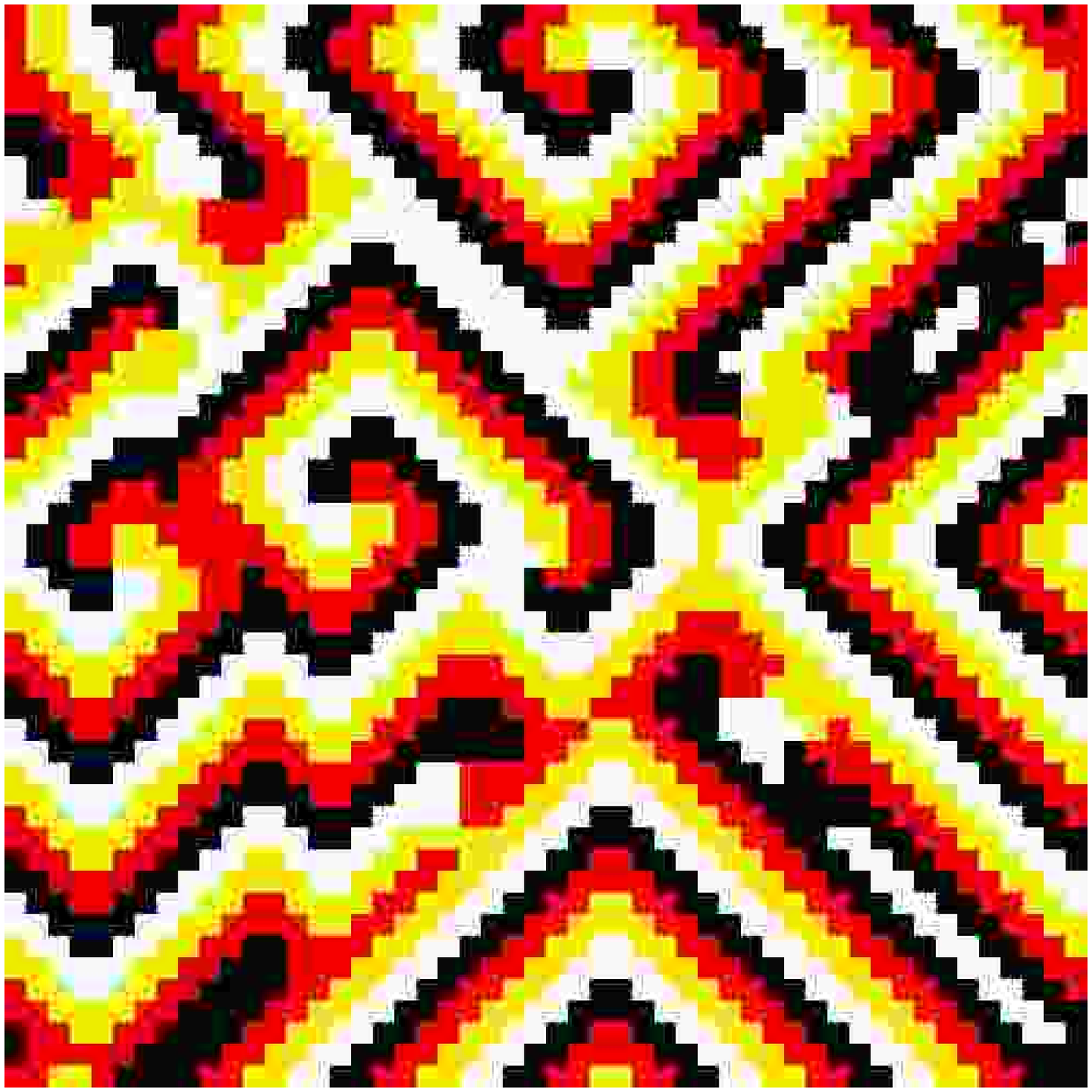}
  \hfill\mbox{}
  \caption{Local sensitivity ($a$) for the cyclic cellular automaton,
    calculated with $p=1,\ f=1$, and corresponding configuration ($b$).}
  \label{fig:cyclic-lle-lpr1}
\end{figure}

Figure \ref{fig:cyclic-lle-lpr0} shows the local sensitivity field for the same
configuration as in Figure \ref{fig:CCA-screenshot}, with a one-cell
perturbation range ($p=0$).  Figure \ref{fig:cyclic-lle-lpr1}, on a smaller
field, shows that with $p=1$ we get qualitatively similar results.  The spiral
cores are easily spotted in both figures, demonstrating that they are
autonomous objects.  There is no clear difference between spiral angles (the
horizontal and vertical lines) and spiral domains.  Spiral boundaries are white
(lowest sensitivity), because any perturbation here will be quickly erased
under the pressure of the radiating spiral cores
(cf.\ \cite{Winfree-time-breaks-down}).  Calculating local sensitivity for 2+1D
automata is very slow: each cell has 8 neighbors and there are 4 states, which
makes $4^9 -1$ possible perturbations.  While sampling on a random subset of
all possible perturbations would be faster, the resulting approximation error
would have to be carefully determined.

We show the CCA field filtered with respect to statistical complexity in Figure
\ref{fig:C-vs-F-fields}.  Comparison with Figure \ref{fig:free_energy} shows
that causal states and order parameter analysis yields to the same information.
The spiral cores are still among the most complex areas.  Here spiral
boundaries are also complex: under statistical complexity filtering, this is
because some prediction is possible, but requires much information; under free
energy analysis, this is due to high phase differences.

\subsection{Comparison of Free Energy, Sensitivity and Complexity}

The three methods described here---the free energy, the local sensitivity and
the statistical complexity---all uncover significant structures in the
spatiotemporal dynamics of the CA.  However, the methods are not equivalent,
and emphasize different aspects of those structures, as one might expect of
different filters.

\begin{figure}[thbp]
\centering
\includegraphics[width=\columnwidth]{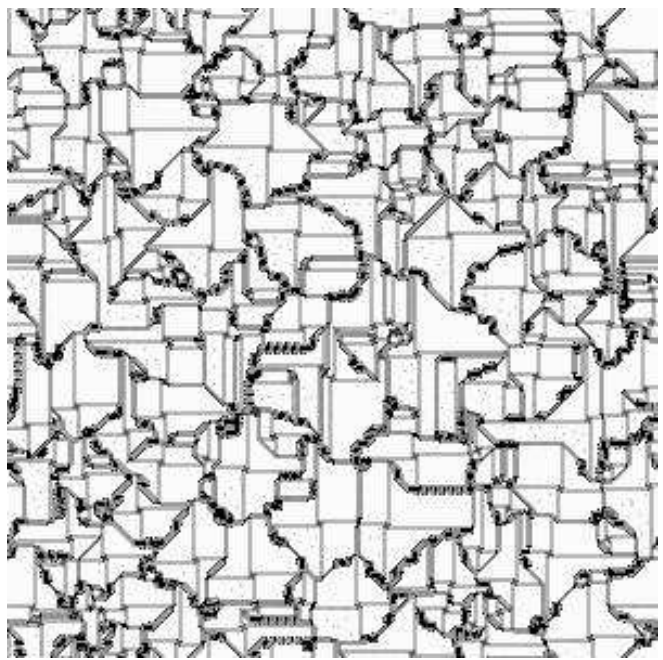}
\caption{Local statistical complexity of the CCA configuration field in Figure
  \ref{fig:CCA-screenshot}, calculated with depth 1 light-cones.  Darker cells
  have higher values for the field.  This figure should be compared to both
  Figures \ref{fig:free_energy} and \ref{fig:cyclic-lle-lpr0}.  Note that free
  energy and statistical complexity are strongly correlated, while local
  sensitivity and statistical complexity are not because they emphasize
  different aspects of the coherent structure.}
\label{fig:C-vs-F-fields}
\end{figure}

Qualitatively, we can see the difference in the order in which different
features are ranked.  For sensitivity, the ordering is particles (spiral
cores), domains, and then domain walls last.  For both complexity and free
energy, the order is by contrast domain walls, particles and domains.  A
quantitative calculation of the correlation coefficients between the different
filtered fields confirms this qualitative impression.  The most strongly
correlated---as one might expect by comparing Figure \ref{fig:free_energy} with
Figure \ref{fig:C-vs-F-fields}---are the statistical complexity and the free
energy, $\rho = 0.690$; we discuss the roots of this strong correlation below.
The complexity and the sensitivity, however, are almost completely
uncorrelated, just as we observed in the case of rule 110---the correlation
coefficient calculated from sample data, $\rho = -0.006$, is not materially
different from 0.  Unsurprisingly, there is also no important correlation
between the sensitivity and the free energy ($\rho = -0.008$).  Despite the
fact that sensitivity and complexity are nearly orthogonal, plotting complexity
as a function of sensitivity (not shown) reveals an interesting relationship:
as sensitivity increases, the {\em minimum} value of complexity rises, though
the converse is not true (minimally sensitive points are found at all values of
complexity).

Sensitivity looks for regions that are unstable to perturbation; because of
this instability, prediction requires very precise discriminations among
histories.  Moreover, localized unstable regions are presumably rare, and the
corresponding states uncommon.  (Recall that the complexity is both
$H[\LocalState]$ and $I[\LocalState;\Past]$.)  Thus the spiral cores are rare,
sensitive and complex.  On the other hand, uncommon, complex structures need
not be unstable.  Domain walls, for instance, are comparatively rare, and
require considerable historical information for their prediction, essentially
because they are regions where the evolution of the phase is hard to determine;
this in turns means that their causal states need more bits to specify.  They
are, however, very stable, because they have considerable spatial extent, and
to destroy or move a wall implies changing the domains on either side.  The
domains themselves, while only minimally complex, are more sensitive than the
domain walls, because perturbations there can create {\em localized} defects.

The relationship between free energy and complexity, while strong and at first
sight clear, is actually a bit tricky.  The easy, but flawed, argument for why
effective free energy and complexity should be correlated runs as follows.  The
effective free energy relies on the assumption that the microscopic dynamics
tend to create ordered regions, where ``ordered'' is understood in the sense
given by the order parameter.  All departures from the ordered state are
unlikely, and large departures are exponentially unlikely.  (This is not
tautologically true, but when it is, we can define an effective free energy.)
These departures take the form of topological defects, like spiral cores and
domain walls, which pay an energetic penalty.  The exact size of this penalty,
and the resulting rarity of these features, will depend on the details of the
microscopic dynamics.  But, precisely because they are rare, the causal states
corresponding to them should be rare too, and so have high local complexity.

The flaw is that the causal state at a point is a function of the configuration
in its {\em past} light cone, while the order parameter and the free energy are
functions of the {\em immediate} neighborhood configuration.  Since the two
filters use different inputs, their strong correlation is not trivial.
However, under fairly general conditions, the local causal states form a Markov
random field, and the Gibbs-Markov theorem tells us that a random field is
Markovian if and only if there is an additive effective free energy which gives
the probability of configurations \cite{Griffeath-on-random-fields}.
Conversely, it can be argued that a complete set of thermodynamic
macro-variables should be equivalent to the causal state of the system, and
should be Markovian in time, so that they can be calculated entirely from the
present configuration \cite{What-is-a-macrostate}.  Thus, there should, in
fact, be a relationship between the complexity and the free energy, {\em if} we
have identified the proper order parameter; the strong correlation we find
between the two fields suggests that we have done so.  The fact that the
correlation, while strong, is not perfect could be either due to our definition
of the order parameter being slightly off, or due to finite-sample errors in
the identification of the causal states.\footnote{Eric Smith (personal
  communication) has pointed out that this argument suggests a relationship
  between the time depth needed to identify the causal states, and the spatial
  range needed to calculate the free energy.  However, pinning down that
  relationship would require careful treatment of many mathematical issues
  regarding sofic shifts, long-range correlations in Markovian fields, etc.,
  beyond the scope of this paper.}

\section{Conclusions}

We have introduced two complementary filtering methods for spatially extended
dynamical systems.  Local sensitivity calculates the degree to which local
perturbations alter the system, and picks out autonomous features.  Local
statistical complexity calculates the amount of historical information required
for optimal prediction and identifies the most highly organized features.  We
emphasize that ``organized'' is not equivalent to ordered, because the latter
corresponds to low entropy which is not the same as high complexity or
organization \cite{QSO-in-PRL}.  A regular lattice is highly ordered, but not
very organized.  On the other hand, high complexity is not equivalent to high
entropy.  A random field has high entropy but low complexity.  Complexity, as
we said earlier, lies between order and randomness, and is a reflection of the
probabilistic properties of the system.  In contrast, sensitivity is a measure
of the system's dynamical properties.  The two are linked through ergodic
theory, but they remain distinct.  Both sensitivity and complexity pick out
spatiotemporal coherent structures, and the structures they identify match
those known from previous, more {\em ad hoc} approaches, whether based on
regular languages (as in ECA) or order parameters and topological
considerations (as in CCA).  In no case, however, is prior knowledge about the
system or its coherent structures used in constructing our filters; at most we
have tuned calculational parameters, in a way akin to adjusting a microscope
until the image comes into focus.  Ideally, one would use both sensitivity and
complexity filtering, because they provide distinct kinds of information about
the system, but we suspect complexity will be much easier to calculate from
empirical data since it only requires observation and not perturbation.

Many theoretical and mathematical questions present themselves about both
filtering methods.  We have mentioned some in passing; here we wish to
highlight just a few.  On the local sensitivity method: What is the exact
relationship between the perturbation range $p$ and the spatial scale of
identified structures?  How much error would be introduced by averaging over a
random subset of perturbations, rather than an exhaustive enumeration?  Can one
identify a typical lifespan for a perturbation, after which it is erased by its
surroundings, and if so is this lifespan related to $f$?  (This last is
presumably related to dynamical mixing properties.)  On the local complexity
method: what quantitative factors relate the volume of data available to the
error in our estimates of the causal states and so of the complexity? Can we
use causal states to give algebraic/automata-theoretic definitions of
``domain'' and ``particle'', like those in the 1D case, without entangling
ourselves in the difficulties of higher-dimensional languages?
(Cf.\ \cite[\S10.5]{CRS-thesis}.)  When we do have partial knowledge of the
correct pattern basis, who can we use this to hasten the identification of the
local causal states?  Perhaps most ambitiously, could one reverse engineer a
good order parameter from the local causal state field and its transition
structure?

As very large, high-dimensional data sets become increasingly common and
important in science, human perception will become increasingly inadequate to
the task of identifying appropriate patterns
\cite{Hand-Mannila-Smyth,Young-summarizing-complexity}.  It is desirable,
therefore, to move towards more automatic filtering methods, and automatic ways
of detecting coherent objects.  Because our filtering methods do not presuppose
any prior knowledge or require human insight about what the right structures
are, they should work generically, across systems of highly varying nature and
dynamics.  This is a hypothesis rather than a theorem, but it can be tested
simply by applying our filtering methods to a wide range of systems with known
coherent structures---and to others where the appropriate structures are {\em
  not} known.  In the latter cases, the test of our methods will be whether the
structures they identify can be used to frame interesting and insightful
higher-level models about the dynamics and functions of the systems involved
(cf.\ \cite{What-is-a-macrostate,%
  Israeli-Goldenfeld-irreducibility-and-predictability}).

{\em Acknowledgments} Thanks to Michel Morvan for valuable discussions, and for
arranging JBR's visit to Ann Arbor; to Vincenzo Capasso for References
\cite{Kolmogorov-light-cones} and
\cite{Capasso-Micheletti-spatial-birth-growth}; to Scott Page and Anna
Amirdjanova for their generous construals of ``diversity'' and ``filtering'',
respectively; and to Dave Albers, Sandra Chapman, Tracy Conrad, Chris Genovese,
David Griffeath, J. Hines, Alfred H{\"u}bler, J{\"u}rgen Jost, Kara Kedi, Eric
Mjolsness, Eric Smith, Padhraic Smyth, Peter Stadler, Naftali Tishby, Dowman
Varn and Nicholas Watkins for comments, discussion and advice.

\appendix

\section{Attempts to Adapt Lyapunov Exponents to Cellular Automata}
\label{app:lyap_review}

There have been at least three attempts to define Lyapunov exponents for
cellular automata.  We review them in order of priority, concluding that none
of them is altogether suited to the aims of this paper.  All are based on
measuring the rate at which small perturbations cause a divergence from the
original trajectory.

Shereshevsky \cite{Shereshevsky-CA-Lypapunov} defines a Lyapunov exponent for
one-dimensional cellular automata, as the velocity of propagation of the edge
of the difference plume.  More exactly, he defines two Lyapunov exponents, for
the velocities of the left and right edges.  By assuming an invariant measure
over configurations exists and has certain properties, he is able to show that
these velocities have reasonable long-time limits.  (Tisseur
\cite{Tisseur-CA-and-LEs} shows the limit exists with somewhat weaker
<conditions on the invariant measure.)  To show that the limit is uniform over
the lattice, he invokes a further spatial ergodicity property.

We, of course, would like things \emph{not} to be uniform, so the fact that we
don't have measures which are so nicely ergodic and stationary should not
trouble us.  But this definition allows a cell which only makes a difference to
one other cell in the future to count as highly influential, provided that said
cell moves very rapidly.  (Consider the shift rule 170 ``copy your neighbor to
your left''.)  Looking at the area of the difference plume (i.e., the number of
differing cells between the original configuration and the perturbed
configuration) seems far more reasonable.

Bagnoli et al. \cite{Bagnoli-et-al-Lyapunov-exponents} come closest to the way
we define local sensitivity in section \ref{sec:lle}, by looking at the total
number of ``defects'', here meaning the number of cells which are different
between the original and the perturbed configuration.  They proceed as follows.
They perturb a single site of the lattice, and iterate forward one time step,
so that the the difference plume now embraces $m$ sites.  They then create $m$
copies of the lattice, each identical to the time-evolved unperturbed
configuration, except at one of the $m$ sites in the difference plume.  They
then repeat this procedure, accumulating more and more copies of the lattice,
each of which differs from the unperturbed trajectory only at a single site.
Suppose the initial perturbation was applied at cell $\vec{x}_0$ and time
$t_0$.  Then for each site $\vec{x}$ and time $t \geq t_0$, the number of
copies of the system which have a defect at $\vec{x}$ at time $t$ is
$N_{\vec{x}_0,t_0}(\vec{x},t)$, and $N_{\vec{x}_0,t_0}(t) =
\sum_{x}{N(\vec{x},t)}$.  $N_{\vec{x}_0,t_0}(t)$ can grow exponentially, and
Bagnoli et al. define the local Lyapunov exponent to be that exponential growth
rate.

This is far from anything that could be called a Lyapunov exponent in the
strict sense of the term.  Furthermore, to avoid the extremely involved
(exponential) calculation which their definition implies, they make use of a
kind of derivative, as in the continuous-system definition of the Lyapunov
exponent.  But these derivative are only defined for Boolean-valued rules,
which makes them useless for, e.g., cyclic cellular automata.  The direct
calculation could be somewhat simplified for deterministic cellular automata by
means of dynamic programming, but it would still be a hard calculation to
obtain a number whose physical significance is unclear.

Finally, Urias et al. \cite{Urias-et-al-sensitive-dependence} define a quantity
which is a Lyapunov exponent (and doesn't require binary values), but only
under a very specific metric, with no clear physical interpretation; it is not
clear how, if at all, their metric could be extended to higher-dimensional
cellular automata.  The ultimate result they get is that the Lyapunov exponent
is just the maximum velocity at which the envelope of the difference region
spreads---taking the maximum over all possible semi-infinite (not single-point)
perturbations.

In this paper we wished not to measure a single Lyapunov exponent for the
entire system, but rather the local (in both space and time) effects of
perturbations.  This allowed us to determine the degree to which different local
structures of the CA are autonomous.

\section{Order Parameter and Free Energy for Spiral-Forming Cyclic Cellular Automata}
\label{app:order-parameter}

A classic method of describing the equilibrium ordered phases of a system is to
define an appropriate order parameter for each phase and an associated free
energy determined from symmetry considerations.  In this appendix we will
perform this analysis for the spiral cellular automaton in its equilibrium
state.  By ``equilibrium'' we mean the cellular automaton's long time behavior
after the domain walls and spiral cores have stabilized.  The transient period
of the cellular automaton corresponds to the phase transition which is not
described by the order parameter formalism.  Examination of the cellular
automaton field (e.g., Figure \ref{fig:CCA-screenshot}) reveals that, while it
certain possesses order, it is also highly frustrated and exhibits many
topological defects, most notably the vortices at the spiral centers and the
domain walls between adjacent spirals.  However the order parameter is defined,
we expect the free energy to increase at these defects.

We begin by defining an appropriate order parameter on the discrete lattice.
We take our cue from the fact that for a continuous field, the spirals would
consist of concentric rings possessing two dimensional rotational symmetry.  We
therefore construct a discretized $XY$ model order parameter and free energy.
In the $XY$ model the Ginsburg-Landau free energy density at a given lattice
site is:
\begin{eqnarray}
F(x,y) = \alpha (\nabla\Phi(x,y))^2
\end{eqnarray}
$\alpha$ is a positive constant which we arbitrarily set to one and henceforth
ignore.  $\Phi$ is a localized phase defined at each lattice site.  In practice
we calculate the Laplacian using each lattice site's nearest neighbors.
\begin{eqnarray}
F(x,y)&=&\frac{1}{4} \lbrace 
(\Psi(x+1,y) - \Psi(x,y))^2  \nonumber\\ &+&
  (\Psi(x,y) - \Psi(x-1,y))^2  \nonumber\\
&+&  (\Psi(x,y+1) - \Psi(x,y))^2  \nonumber\\ 
&+&  (\Psi(x,y) - \Psi(x,y-1))^2 \rbrace
\end{eqnarray}
$\Phi$ is defined in terms of the states of the configuration field $\sigma \in
\{0,1,2,3\}$.  Noting that the spiral cellular automaton field consists mainly
of plane waves cycling through the four colors and traveling mainly along the
diagonals, although at times along the x and y axes, it seems reasonable to
define the spiral cellular automaton phase $\Phi$ as a function of the {\it
  direction} in which the wave is traveling.  We define a normalized two
component wave vector at each lattice site $(x,y)$
\begin{eqnarray}
{\bf \Psi}(x,y) &=& \frac{1}{\sqrt{\Psi_x^2 + \Psi_y^2}}\big(\Psi_x(x,y)\hat x + \Psi_y(x,y)\hat y\big) \nonumber\\
 &=& (\bar\Psi_x,\bar\Psi_y)
\end{eqnarray}
where $\bar\Psi_x$ and $\bar\Psi_y$ are the components of the normalized vector
and the unnormalized components $\Psi_x$ and $\Psi_y$ are defined as follows.
\begin{eqnarray}
\Psi_x(x,y) &=& \sum_{i=\pm 1} g(\sigma(x+i,y),\sigma(x,y))\\
\nonumber
& & + \frac{1}{2}\sum_{i=\pm 1, j=\pm 1} g(\sigma(x+i,y+j),\sigma(x,y))
\end{eqnarray}

\begin{eqnarray}
\Psi_y(x,y) &=& \sum_{i=\pm 1} g(\sigma(x,y+i),\sigma(x,y))\\
\nonumber
& & + \frac{1}{2}\sum_{i=\pm 1, j=\pm 1} g(\sigma(x+i,y+j),\sigma(x,y))
\end{eqnarray}
Note that the first sum is over nearest neighbors in the $x$ (resp. $y$)
direction and the second sum is over all next nearest neighbors.  The factor
$1/2$ comes from the corresponding $1/2$ in the definition of the order
parameter.  The function $g$ is defined as
\begin{eqnarray}
g(\sigma_1,\sigma_2) &=& +1 \; \; \; \mathrm{if} ~ \sigma_1 = \sigma_2 +1 \nonumber\\
&=& -1 \; \; \; \mathrm{if} ~ \sigma_1 = \sigma_2 - 1 \nonumber\\
&=& 0 \; \; \; \; \; \mathrm{otherwise}
\end{eqnarray}
It is to be understood that the above definition of g involves addition and
subtraction modulo 4.

The phase $\Phi$ at a given lattice site is defined simply as
\begin{eqnarray}
\Phi = \tan^{-1}\frac{\bar\Psi_y}{\bar\Psi_x}
\end{eqnarray}
Thus we see that in the upper right quadrant of a spiral,
$\Psi=(\frac{1}{\sqrt{2}},\frac{1}{\sqrt{2}})$ and $\Phi=\frac{\pi}{4}$,
whereas in the upper left quadrant $\Psi =
(\frac{1}{\sqrt{2}},-\frac{1}{\sqrt{2}})$ and $\Phi = \frac{3\pi}{4}$.  Thus
there is a phase gradient along the boundary between these two quadrants of the
spiral and an increase of free energy here.  This can be seen in Figure
\ref{fig:free_energy}.  (The localization of the free energy increase to the
boundary between quadrants is, of course, an artifact of the discretized
spatial lattice.  In the continuum limit, the gradient energy would be
distributed evenly around the central vortex.)  The other locations where one
expects to see increased free energy are in the vortex cores, and at the
boundaries between spirals.  Our free energy captures both of these effects
(Figure \ref{fig:free_energy}).

\bibliographystyle{apsrev}
\bibliography{locusts}

\end{document}